# "Without in Any Way Disturbing the System": Illuminating the Issue of Quantum Nonlocality

## (A Response to Andrei Khrennikov)

## Arkady Plotnitsky*

*If, without in any way disturbing a system, we can predict with certainty (i.e., with probability equal to unity) the value of a physical quantity, then there exists an element of physical reality corresponding to this physical quantity.*
—A. Einstein, B. Podolsky, and N. Rosen, "Can Quantum-Mechanical Description of Physical Reality be Considered Complete?" (Einstein et al 1935, p. 138)

**…** not being any longer in a position to speak of the autonomous behavior of a physical object, due to the unavoidable interaction between the object and the measuring instruments.
—N. Bohr, "Complementarity and Causality" (Bohr 1937, p. 79)

Any "entanglement of predictions" that takes place can obviously only go back to the fact that the two bodies at some earlier time formed in a true sense *one* system, that is, were interacting, and have left behind traces of each other. If two separate bodies, each by itself known maximally, enter a situation in which they influence each other, and separate again, then there occurs regularly that which I just called *entanglement* of our knowledge of the two bodies.
—E. Schrödinger, "The Present Situation in Quantum Mechanics" (Schrödinger 1935, p. 161)

**Abstract**. In several recent communications (Khrennikov 2019a, b, c, 2020a, b), A. Khrennikov argued for "eliminating the issue of quantum nonlocality" from the analysis of quantum entanglement and quantum phenomena in general. He proposed to differentiate quantum and classical phenomena and entanglement not by their respective nonlocality and locality, as is common, but by the discreteness of quantum phenomena vs. the continuity of classical phenomena, supplemented by Bohr's complementarity in the case of quantum phenomena. As I argue here, however, the question may not be that of "*eliminating* the issue of quantum nonlocality" but instead of *illuminating* this issue, a task that, I also argue, can be pursued by relating quantum nonlocality to other key features of quantum phenomena. I suggest that the following features of quantum phenomena and quantum mechanics, distinguishing them from classical phenomena and classical physics—(1) the irreducible role of measuring instruments in defining quantum phenomena, (2) discreteness, (3) complementarity, (4) entanglement, (5) quantum nonlocality, and (6) the irreducibly probabilistic nature of quantum predictions—are interconnected in defining quantum phenomena and distinguishing them from classical ones, so that it is difficult to give an unconditional priority to any one of them. To argue this case, I consider quantum phenomena and quantum mechanics from a nonrealist or, in terms adopted here, "reality-without-realism" (RWR) perspective. This perspective extends and gives new dimensions to Bohr's view, grounded in his analysis of the irreducible role of measuring instruments in the constitution of quantum phenomena, with the quantum measurement defined by the entanglement between the quantum object under investigation and the instrument used.

## 1. Introduction

In several recent communications, A. Khrennikov argued for "eliminating" the considerations of nonlocality from the analysis of quantum entanglement and, by implication, quantum phenomena in general (Khrennikov 2019a, b, c, 2020a, b).[1] In (Khrennikov 2019b, 2020b), he proposed to differentiate classical and quantum entanglement not by their respective locality and nonlocality, as has been common in discussions of quantum entanglement, but instead by the inherent discreteness of quantum phenomena vs. the inherent continuity of classical phenomena.[2] It would be more accurate to speak of the continuity of the processes underlying and connecting classical phenomena, given that some classical phenomena are observationally discrete. By the same token, there is also a question whether this type of continuous connectivity, commonly assumed in classical physics or relativity (that is, the idealization of nature that either theory provides), is also possible to assume, at least ideally and in principle, in the case of quantum phenomena. This has been one of the main foundational questions of quantum theory from its inception on, and it is at stake in Khrennikov's papers here cited and in this paper.

* Theory, Literature, and Cultural Studies Program, College of Liberal Arts, Purdue University, W. Lafayette, IN, 47907, email: plotnits@purdue.edu, https://orcid.org/0000-0001-6462-2181

Khrennikov's argument offers a useful angle on quantum phenomena and entanglement, first of all, by directing our attention to the difficulties and sometimes confusion surrounding the concept of quantum nonlocality, which part of his argument is not uncommon, and, secondly, by emphasizing the discreteness of quantum phenomena and, with it, the role of Bohr's concepts of complementarity and phenomenon in this problematic, which is far less common.[3] The discreteness of quantum phenomena has often been neglected or taken for granted in recent discussions of entanglement or quantum phenomena in general. While complementarity is occasionally brought up in these discussions, Bohr's concept of phenomena is disregarded nearly altogether, except for treatments offered by scholars of Bohr, the present author among them (e.g., Plotnitsky 2009, pp. 327-336; 2012, pp. 138-149; 2016, pp. 155-168). The role of quantum discreteness was more prominent at earlier stages of the development of quantum theory and the debate concerning it, beginning with M. Planck's discovery of the discrete nature of radiation in certain circumstances (previously assumed to be continuous phenomena in all circumstances) and, especially, following Bohr's 1913 atomic theory (Bohr 1913), and then the discovery of quantum mechanics (QM) in 1925 and Bohr's interpretation of quantum phenomena and QM, via his concept of complementarity. This interpretation was introduced in 1927 in the so-called Como lecture, "The Quantum Postulate and the Recent Development of Atomic Theory," grounded in what Bohr called "the quantum postulate, which attributed to any atomic process an essential discontinuity, or rather individuality, completely foreign to the classical theory and symbolized by Planck's quantum of action [$h$]" (Bohr 1987, v. 1, p. 53).

In fact, discontinuity and individuality are equally essential, because each observed quantum phenomenon is both individual, even unique, in itself and is discontinuous with, discrete in relation to, any other quantum phenomenon. The phrase "atomic process" leaves some space for ambiguity concerning its meaning, as do several other aspects of Bohr's argument in the Como lecture. Bohr was never satisfied with his argument there and revised it, in part under the impact of his initial exchanges with A. Einstein on the epistemology of quantum theory in October 1927, on several key points, in particular by removing certain remnants of realism and causality (classically defined) still found there, even before the lecture, delivered in Como in September 1927, was published in April 1928 (Bohr 1928).[4] Eventually, sometime around 1937, following his exchanges with Einstein concerning the thought experiment designed by Einstein, B. Podolsky, and N. Rosen, the EPR experiment (Einstein et al 1935), Bohr introduced his concept of "phenomenon," in which, along with complementarity (central to all versions of his interpretation), he grounded by and large the final version of his interpretation, with only a few minor revisions added subsequently.[5]

---

[1] The most recent papers (Khrennikov 2020a, b) also respond to the first version of the present paper (Plotnitsky 2019d), and (Khrennikov 2020b) is a revised and extended version of (Khrennikov 2019b), addressed in the first version. Apart from the Appendix, however, this version is neither a counter-response to Khrennikov's comments on the first version nor a new response to his overall argumentation in these papers. It only aims to provide additional details to my argument without essentially changing it.
[2] Given that in this paper I will only be concerned with quantum entanglement, the term entanglement will from now on refer to quantum entanglement.
[3] The subject has been addressed, in considering Bohr's interpretation and the Bohr-Einstein debate (rather thanalong the lines of Khrennikov's argument in question), in previous works by the present author (Plotnitsky 2009, pp. 237-312; 2012, pp. 107-136; 2016, pp. 136-154, 2019c), which Khrennikov cites.

In Bohr's definition, a phenomenon in quantum physics is defined by what is observed, in the specified measuring arrangement of each experiment, as the result of the interaction between the quantum object under investigation and the measuring instrument, prepared in accordance with this specification, to be used (e.g., Bohr 1987, v. 2, p. 64). This concept helped Bohr to emphasize the defining role of measuring instruments in the constitution of quantum (vs. classical) phenomena, a role central to his argumentation all along, and to sharpen the point that the discreteness and individuality in question are those of quantum *phenomena*, observed in measuring instruments, rather than the Democritean atomic discreteness of quantum *objects*, with which our instruments interact (Bohr 1987, v. 2, pp. 32-33).[6] In Bohr's ultimate interpretation, because of the irreducible role of measuring instruments in the constitution of all quantum phenomena, one can, as against classical physics, no longer describe the independent behavior of quantum objects at all (either as continuous or discrete), by extracting them from the indivisible enclosures of phenomena. This impossibility defines what I call the "reality-without-realism" (RWR) view and the corresponding interpretations (there could be more than one) of quantum phenomena. This view, however, only applies to quantum *objects*, or the stratum of reality this concept idealizes, by placing this stratum beyond representation or even conception, rather than quantum *phenomena*, in which case a representation and, thus, realism are possible. Realism, by contrast, is defined by the assumption of the possibility of such a representation or at least conception. In RWR-type interpretations, beginning with that of Bohr, each quantum phenomenon is always discrete in relation to any other quantum phenomenon, without allowing us to know or possibly even conceive of how the quantum phenomena come about, specifically by means of a continuous and causal process that defines classical physics or relativity. Quantum phenomena may, however, be related to each other, by means of QM, or some other theory, in terms of probabilistic or statistical predictions or correlations. No other predictions or correlations are possible on experimental grounds, at least as things stand now, because the repetition of identically prepared quantum experiments in general leads to different outcomes.

As indicated by the more moderate subtitles of (Khrennikov 2019b, 2020b), "eliminating *the issue* of quantum nonlocality," paraphrased in my subtitle, Khrennikov does not appear to entirely deny that there is quantum nonlocality or something that could be so named. Instead, he appears to suggest that, whatever quantum nonlocality may be, it is not defining or even important, *is not the issue*, in considering quantum phenomena, including those of entanglement.[7] He also argues, in my view, rightly, that QM and quantum phenomena are "local" insofar as they do not entail any instantaneous transmission of physical influences between spatially separated physical systems or "a spooky action at a distance" [*spukhafte Fernwirkung*], famously invoked by Einstein in this connection (Born 2005, p. 155). Such an action would bring QM in conflict with relativity, which prohibits physical influences propagating faster than the speed of light in the vacuum, $c$. In his reply to EPR, Bohr, by contrast, argued for the compatibility of quantum phenomena and QM, at least in his interpretation, for "all exigencies of relativity theory" and thus for its locality in the sense of the absence of an action at a distance (Bohr 1935, p. 701n). Although adopted by some, the idea that QM is nonlocal in this sense and the corresponding concept of "quantum nonlocality" could indeed be "misleading," as Khrennikov suggests (Khrennikov 2020b, p. 15).

---

[4] Bohr dated the lecture 1927 when it was republished in his book, *Atomic Theory and the Description of Nature* (Bohr 1934), now republished as the first volume of (Bohr 1987).

[5] I have considered different versions of Bohr's interpretation in detail in (Plotnitsky 2012). It is worth noting in this connection that there is no *single* Copenhagen interpretation, given that even Bohr changed his views, sometimes significantly, a few times, in part (there were other factors) under the impact of his life-long debate with Einstein. It is, accordingly, more fitting to speak, as W. Heisenberg did of "the Copenhagen spirit of the quantum theory" (Heisenberg 1930, p. iv). This spirit pervades a spectrum of interpretations of quantum phenomena and QM which share some of their features, most especially the irreducible role of probability in QM, but not all of them, which is, again, true as concerns Bohr's several interpretations, the first of which retained realism and classical causality.

[6] Around 1937-1938, Bohr also introduced, alongside his concept of phenomena, a new concept of "atomicity," pertaining strictly to quantum phenomena rather than quantum objects (e.g., Plotnitsky 2012, pp. 138-150). For all practical purposes, however, this concept is equivalent to Bohr's concept of phenomenon and only highlights some of the aspects of the latter concept.

On the other hand, Einstein himself did not define this nonlocality as "quantum nonlocality," the term introduced later in the wake of Bell's theorem. He only argued, most famously in EPR's paper, that QM is either incomplete or nonlocal in allowing for an instantaneous action at a distance (Einstein et al 1935). The concept of completeness used by Einstein requires further qualifications to be offered below. For the moment, I shall call this nonlocality "the Einstein-nonlocality," as against "quantum nonlocality," which can and has been defined differently. I shall now offer one such definition. As Khrennikov appears to admit (while denying that QM needs to be seen as Einstein-nonlocal), one might in the case of quantum phenomena speak of, in terms of the present author, "spooky *predictions* at a distance," without necessarily assuming any spooky *action* at a distance (e.g., Plotnitsky 2009, 2012, 2016, 2019b).[8]

The reason for adopting this view is as follows. There are, experimentally confirmed, statistical correlations between certain, specifically prepared, arbitrarily distant quantum events, which correlations are also properly predicted by QM (both of which facts are accepted by Khrennikov). These predictions are "spooky" insofar as, against classical physics or relativity, there is, at least in RWR-type interpretations, no story to be told and no concept to be formed of how these correlations or quantum phenomena, in the first place, come about or why these predictions are possible. At the same time, neither these correlations themselves nor predicting them need entail "a spooky *action* at a distance" or Einstein-nonlocality, including those of the EPR-type, in which case certain predictions between such events are even possible with the probability one, albeit with certain crucial qualifications, explained below. I shall, then, define the concept of "quantum nonlocality" as referring to the existence of these correlations between *distant* quantum events and the possibility of predicting these correlations or probabilities or statistics of quantum events, thus, predicting them at a distance. Indeed, as I argue here, all quantum predictions are predictions at a distance, without, again, necessarily implying an action at a distance, although they may and have been interpreted in this way.

---

[7] Khrennikov appears to take a more uncompromising position against quantum nonlocality in (Khrennikov 2020a), thus possibly implying that (Khrennikov 2019a, 2020b) adopt a stronger position in this regard as well, which is why I qualify my assessment by "appear." Even if this is the case, my main argument here, which advocates a certain specific conception of quantum nonlocality, is not affected. I shall, however, comment on Khrennikov's argument in (Khrennikov 2020a) in the Appendix.

[8] I should qualify that, while Khrennikov credits the concept of "predictions at a distance" as valid and while he does not appear to deny the possibility that these predictions might be "spooky" insofar as there is no realist and specifically causal description of how these predictions come about, he prefers to avoid the "spookiness" of these predictions (Khrennikov 2020a, pp. 15-17). I shall return to this subject in the Appendix. My choice of the word "spooky" is motivated by Einstein's use of it in his famous phrase. One might speak more neutrally of enigmatic or unexplainable predictions at a distance, insofar as we don't know or perhaps cannot even conceive either how quantum phenomena come about and, by the same token, why any quantum-mechanical predictions, which are fully in accord with the experimental evidence available thus far, are possible.

Quantum nonlocality is sometimes defined differently, for example, in terms of violations of Bell's or related inequalities, or still other mathematical features, dealing with the data obtained in the corresponding experiments. Such definitions, however, still leave space for their physical interpretation, and quantum nonlocality as just defined provides such an interpretation, *one* such interpretation among other possible interpretations, some of which interpret quantum nonlocality in terms of Einstein-nonlocality. I find the latter interpretations problematic because they are in conflict with relativity. On the other hand, there are interpretations of quantum phenomena and QM, such as those of the RWR-type, beginning with that of Bohr, that only entail quantum nonlocality in the present definition (or other definitions of this type) and avoids Einstein-nonlocality. These interpretations show that Einstein-nonlocality is not necessarily a feature of quantum phenomena or QM, while quantum nonlocality may well be.

Einstein eventually admitted that Einstein-nonlocality (which he meant by "nonlocality") could be avoided if one assumes that QM is a strictly statistical theory that does not provide a representation of, nor predictions concerning, the behavior of the ultimate individual constituents considered. He was, however, not satisfied with this alternative, because it was in conflict with his conviction that a fundamental physical theory should do precisely this. For one thing, why QM was able to make its statistical predictions remained unexplained, as against, for example, classical statistical physics, where the behavior, predictable only statistically, of the systems considered is underlain by the classically causal behavior of their individual constituents, a behavior predictable ideally exactly by classical mechanics. For Einstein, this aspect of QM made the theory akin to magic, "Jacob's pillow," of Göttingen and Copenhagen, and not "the real thing" (e.g., Born 2005, pp. 155, 205, Einstein 1949a, p. 81). "The real thing" could only have been a realist and causal, and even deterministic, theory of the ultimate constitution of nature, just as was relativity, special and general.

Accordingly, it might be that the question is not that of "*eliminating* the issue of quantum nonlocality," but instead that of *illuminating* this issue, even though the ultimate nature of quantum nonlocality itself may remain beyond illumination, beyond any picture or other concept that our thought can form. It is thus beyond any "*Bild*," the concept invoked in this connection, by Khrennikov, via H. Hertz and L. Boltzmann (Khrennikov 2020a). If so, however, it is not because there is a "spooky action at a distance," which would resolve the issue (but, again, not in a way acceptable to either Einstein or Bohr), but because there isn't any such action. In the absence of such an action, it may not be possible to know or even to conceive of the physical reasons for quantum nonlocality or the emergence of quantum phenomena in general, predicting which, I argue, always entails predictions at a distance and thus quantum nonlocality as here defined. That, however, does not mean that the *issue* of quantum nonlocality cannot be further illuminated. One could, I argue, be helped in this task by relating quantum nonlocality more firmly to other key features of quantum physics, discreteness and complementarity among them.

I should offer an immediate disclaimer. I do not imply by this elaboration or by my subtitle that this paper can accomplish this task. Instead, using the sense of "illuminating" as reflecting a continuing process, I hope to contribute to this process, in a positive spirit, rather than with the aim of refuting previous *arguments*, although I will argue against certain previous (and still current) *claims*, beginning with the claim that quantum nonlocality implies Einstein-nonlocality.[9]

With this disclaimer in mind, I shall argue that the following key defining features of quantum phenomena and QM, possibly distinguishing them from classical phenomena and classical physics (there are quite a few of them!)—(1) the irreducible role of measuring instruments in defining quantum phenomena, (2) discreteness, (3) complementarity, (4) entanglement, (5) quantum nonlocality, and (6) the irreducibly probabilistic or statistical nature of quantum predictions, which pertains to our quantum theories rather than quantum phenomena—are all interconnected so that it is difficult to give an unconditional priority to any one of them in interpreting quantum phenomena and QM. In particular, I shall suggest that the discreteness of *quantum phenomena* (which is, again, not the same of the discreteness of *quantum objects*) in fact implies the nonlocality of our predictions concerning them, and thus of QM, at least if one adopts the RWR view of quantum phenomena. Abandoning or even in principle precluding the assumption of the continuous physical connections between quantum events, inevitably makes our predictions concerning them predictions at a distance, without, however, necessarily implying any action at a distance.

I am not saying that it is not, in principle, possible to distinguish quantum and classical phenomena or quantum and classical theory by a single feature, or define them accordingly, as has been suggested in the case of QM, although not quantum phenomena, by recent (reconstruction) projects of deriving QM for discrete, rather than, at least thus far, continuous, variables. Notably, such derivations do not share the same single feature distinguishing quantum and classical theories (which theories still shares other features, as is the case in classical and quantum mechanics).[10] For the reasons explained below, it is tempting to argue, following Bohr, that, if there is any single feature distinguishing classical and quantum physics, it is the irreducible role of measuring instruments in defining quantum phenomena, which makes it very difficult and may even preclude (it does not in RWR-type interpretations) representing the independent behavior of quantum objects, as against their effects on measuring instruments. This feature would equally apply to discrete and continuous quantum variables. One might, however, prefer to err on the side of caution. For example, while discreteness is automatic under this assumption, complementarity is not. Also, although an entanglement is part of the interaction between quantum objects and measuring instruments (because this interaction entangles them), and although these interactions ground quantum nonlocality, both entanglement and quantum nonlocality are more general features of quantum phenomena, rather than merely consequences of these interactions. This greater generality is manifested in the EPR-type experiments, because there we deal with two entangled quantum objects, even though, as Bohr argued, the irreducible role of measuring instruments remains part of any EPR-type experiment, a role, he also argued, underappreciated by EPR and Einstein in his subsequent communications (Bohr 1935). Accordingly, it seems more reasonable to focus on all of these features in their interactions, which may still not be exhaustive in defining quantum phenomena vs. classical ones.

---

[9] Quantum nonlocality has been the subject of long-standing and ongoing investigations from diverse and sometimes conflicting perspectives (in part due to different interpretations of quantum phenomena and QM) and has been intensely debated. The literature on the subject is extensive, and my aims and limits here will only allow me to mention a very small portion of it.

[10] A case in point is "the continuity axiom" of L. Hardy's derivation (Hardy 2001) and "the purification postulate" in (D'Ariano et al 2017). The latter is especially relevant in the present context because it is connected to quantum entanglement, and I shall further comment on it below.

For one thing, there is still the role of Planck's constant, $h$. Historically, quantum phenomena were initially defined by the fact that, in considering them, $h$, must be taken into account, which is still the case. Doing so allows one to use classical theory in physically describing observed quantum phenomena (observed in the suitably prepared measuring instruments), although classical physics could not predict these phenomena. This incapacity initially led to the assumption that there must exist entities in nature responsible for these phenomena, entities the behavior of which could be described by classical physics, for otherwise classical physics would be able to predict these phenomena as well. These entities are understood or, in the present view (explained below) are idealized, as quantum *objects,* in contradistinction to quantum *phenomena*, defined, again, by what is observed in measuring instruments and requiring a proper quantum theory to take $h$ into account in order to predict them.[11] While, however, the role of $h$ is physically irreducible in quantum phenomena, their specificity as quantum, again, appears to be defined by a broader set of features, such as those in question in this paper (the role of measuring instruments, discreteness, complementarity, entanglement, and quantum nonlocality), some of which are not physically linked to $h$, at least not expressly. On the other hand, some of these features are also exhibited by classical phenomena or found in alternative theories, such as Bohmian mechanics (which is expressly Einstein-nonlocal) or "toy" models different from those of standard QM.[12] Nevertheless, all quantum phenomena known thus far cannot be meaningfully considered physically or (quantitatively) predicted apart from taking $h$ into account.

It was arguably this fact that compelled Bohr to speak of $h$ as "symbolizing" the essential discontinuity and individuality of quantum phenomena, which express the "essence" of the quantum theory, even if, as I argue here, not all of its essence (Bohr 1987, v. 1, p. 53). As just noted and as will be discussed in detail below, ultimately, for Bohr the main difference between classical and quantum phenomena, is that in considering quantum phenomena we must rigorously differentiate between quantum objects (which, or the reality idealized, are beyond all description, including by the formalism of QM) and measuring instruments (which are described by classical physics) and the impossibility of physically separating the behavior of quantum objects from their interactions with the measuring instruments, which is quantum, finite, and as such is uncontrollable (Bohr 1935, pp. 697, 700). This situation, insofar as we only deal with individual measurements observed strictly in instruments and never, as in classical physics and relativity with the behavior of physical objects that connects these observations, also makes all quantum phenomena discrete in relation to each other. For Bohr, $h$ numerically represents and symbolizes precisely this situation.

Why $h$, as physically the *quantum* of action (in its finite and discrete character and its specific numerical value) is part of nature or, more rigorously, our interactions with nature in the case of quantum phenomena is a separate matter, well beyond QM and, most likely, QFT. As thing stand now, $h$ is just there and must be take into account in all predictions of quantum phenomena. At the same time, from Bohr's or the present perspective, defined by the RWR view, $h$ does not pertain to quantum objects or behavior but only to quantum measurements and our quantum theories, and $h$ is measured classically in the first place. The reason for this is that in the RWR view, the ultimate nature of *reality* responsible for quantum phenomena, commonly seen or (in the RWR view, idealized) in terms of quantum *objects* and behavior, is beyond any representation or even conception, physical or mathematical, which precludes associating any numerical constant with this reality. This is why for Bohr *the essential discontinuity and individuality* defining quantum theory and "completely foreign to the classical theory," based in *the essential continuity* underlying and connecting all classical phenomena, are ultimately due to the irreducible role of measuring instruments in the constitution of quantum phenomena, as "*symbolized* by Planck's quantum of action [$h$]," found in classically described measurements (Bohr 1987, v. 1, p. 53; emphasis added). These continuity and individuality are (regardless of one's interpretation of quantum phenomena and QM) features of our interactions with nature by means of our experimental technology and not necessarily, of the ultimate constitution of nature, to which, in the RWR view, no knowable or even humanly conceivable features can apply.

---

[11] Of course, quantum objects, such as electrons, were initially discovered within the framework of classical physics, and were expected to behave classically, although with the introduction of the photon, a quantum object, to begin with, meeting these expectations became increasingly difficult. Photons are now treated by quantum electrodynamics (QED).

While thus assuming that what is responsible for quantum phenomena exists or is real (and allowing it to be idealized in terms of quantum objects and their behavior), this concept of reality precludes realism, defined by the possibility of representation of the reality considered by the corresponding theory. I shall properly discuss the concept of reality in the next section. Briefly, by *reality* I refer to that which is assumed to exist, without making any assumptions concerning the character of this existence, which allows one to place, as in the RWR view, this character beyond representation or conception. I understand by existence the capacity to have effects on the world with which we interact. Assuming the possibility of representing the character of what is assumed to be real, rather than merely assuming its existence, defines realism.

Importantly, the RWR-type view and the corresponding interpretations of quantum phenomena only assume the RWR view of *the ultimate constitution of reality* responsible for quantum phenomena. This assumption allows that some strata of the overall reality these interpretations consider, which includes that of quantum phenomena, may be represented and thus allow for a realist treatment. In the present interpretation, which, on this point, follows that of Bohr, *the observable parts* of quantum phenomena (the emphasized qualification is, as I shall explain, crucial) are represented by classical physics, while quantum objects and behavior or, again, the reality thus idealized is, as an RWR-type reality, beyond representation or even conception. This difference is subtle, because, while a representation of the reality considered always entails a conception of this reality, the reverse is not necessarily true because a conception of this reality may be short of any workable, especially suitably mathematized, physical representation of it. I shall address this aspect of the situation and possible differences in this regard between Bohr's and the present view in detail below.

I would like to emphasize, in closing this introduction, that all my claims in this paper only concern *interpretations*, those of the RWR type amidst other interpretations (some of which are realist), of quantum phenomena and QM. I make no claims concerning how nature ultimately works. Such claims would, in any event, be precluded by the RWR view, because it places the ultimate working of nature beyond knowledge or even conception, at least as things stand now and possibly ever. One the other hand, the RWR view does allow one to make claims concerning the effects of these workings on those aspects of the world that are available to our thought, representation, and knowledge, such as, with the help of our experimental technology, quantum phenomena observed in measuring instruments. It allows us to make claims concerning our interactions with nature and, as such, *helps* to establish the continuity between quantum physics and the preceding history of physics as a mathematical-experimental science of nature. It is true that one needed new mathematical theories and new experiments to have quantum physics and will continue to need them to advance physics, which obviously requires much more than the RWR view of the ultimate nature of reality. The RWR view is never sufficient for new thinking and knowledge, but it is may be helpful and even necessary for them.

Besides the RWR view also has, at least for some of us, beginning, arguably, with Bohr, a philosophical appeal. As S. Coleman (reportedly) said, in a variation on a more common theme that nature's imagination exceeds our own, that "if thousands of philosophers spent thousands of years searching for the strangest possible thing, they would never find anything as weird as quantum mechanics" (reported Randall 2005, p. 117). The degree to which this is true may be questioned. Philosophers, beginning with the pre-Socratics, or poets, have spent thousands of years exploring things that are pretty weird, or in any event, quantum mechanical-like, as Bohr suggested when he said that "the epistemological situation here [in QM] encountered, which *at least in physics* is of an entirely novel character" (Bohr 1937, p. 87; emphasis added). This means that, at least in some respects, this situation might have been previously encountered elsewhere. In any event, if we bring back realism and (classical) causality to our understanding of the ultimate constitution of nature, or if, finally fulfilling Einstein's hope, fundamental physics return to them, which is possible, the philosophical novelty, defined by "the epistemological situation here encountered, which at least in physics is of an entirely novel character," will be gone from fundamental physics.

On the other hand, it is also conceivable that a physical theory would emerge, perhaps the one bringing gravity and other forces of nature into a harmony, even if not to unify them, that will require a view that is neither realist nor that of RWR-type. According to A. Zeilinger and co-authors: "It may well be that in the future, quantum physics will be superseded by a new theory, but it is likely that this will be much more radical than anything we have today" (Zeilinger et al. 2005, pp. 236–237). Such a theory may be difficult or even, at present, impossible to imagine, but it may come about nevertheless. Quantum phenomena and quantum theory, beginning at least with Bohr's 1913 theory, if not with Planck's discovery of the quantum nature of radiation in 1900, and finally Heisenberg's QM were each pretty unimaginable before they appeared. Indeed, as explained below, relativity, with its new kinematics, already took physics beyond anything previously imagined. Quantum theory was the next step on that road. It may not and, hopefully, will not be not the last one. A return to realism is, again, possible, too, but it would hardly be of much philosophical interest.

The remainder of this paper proceeds as follows. The next section outlines the key concepts and the main argument of this paper. Section 3 considers quantum measurement, as defined by the entanglement between the quantum object under investigation and the measuring instrument used, the situation central to my argument in this paper. One of the surprising conclusions there is that a quantum measurement does not actually measure anything pertaining to the quantum objects under investigation but only the state of the quantum stratum of the instrument uses after the interaction between the instrument and the object. Section 4 discusses complementarity and the concept of quantum causality. Section 5 offers a reassessment of EPR's argument and Bohr's reply, and of the role of quantum nonlocality vs. Einstein-nonlocality in EPR-type experiments.

## 2. "The unavoidable interaction between the object and the measuring instrument": measurement and reality in quantum physics

I would like to begin with a passage from Bohr's important, but rarely cited, 1937 article, "Complementarity and Causality." The article is the first published work by Bohr that presented his ultimate interpretation of quantum phenomena and QM, an interpretation arrived at after a decade of revisions, in part (there were other factor) under the impact of his exchanges with Einstein, especially concerning the EPR experiments, and grounded in the RWR view of quantum objects and behavior, or again, the reality so idealized, as responsible for quantum phenomena. As stated in the Introduction, one feature that most centrally defines the difference between classical and quantum phenomena in all versions of Bohr's interpretation, and in which his ultimate interpretation is based as well, while also giving this feature an RWR-type interpretation, is "the unavoidable interaction" between quantum objects or, again, something in nature so idealized, and measuring instruments. According to Bohr:

> The renunciation of the ideal of causality in atomic physics which has been forced on us is founded logically only on our not being any longer in a position to speak of the autonomous behavior of a physical object, due to the unavoidable interaction between the object and the measuring instrument which in principle cannot be taken into account, if these instruments according to their purpose shall allow the unambiguous use of the concepts necessary for the description of experience. In the last resort an artificial word like "complementarity" which does not belong to our daily concepts serves only briefly to remind us of the epistemological situation here encountered, which at least in physics is of an entirely novel character." (Bohr 1937, p. 87)

Although complementarity does more, it is, as a quantum-theoretical concept (it could be defined more generally and applied elsewhere), essentially connected to this situation and thus to the RWR view, announced by Bohr's claim of "our not being any longer in a position to speak of the autonomous behavior of a physical object." I shall discuss complementarity in Section 4, and shall only note here that it is complementarity that enables this unambiguous use by making some of these concepts (such as those of position and momentum measurements) and the corresponding quantum phenomena, observed in measuring instruments, "complementary": mutually exclusive and yet equally necessary for a comprehensive account of the totality of quantum phenomena.

The concept of causality that grounds the "ideal of causality" in question is defined by the claim that the state of a physical system is determined, independently of observation, in accordance with a law, at any and all future moments of time once it is determined at a given moment of time, and any such determination is itself determined by the same law by the system's previous behavior, by any of its previous states. This assumption, thus, implies a concept of reality first, the reality defining this law and making the concept of causality ontological (pertaining to the nature of reality). This concept of causality has a long history, beginning with the pre-Socratics, and it has been effective in classical physics, as part of the concept of reality assumed there. I shall term this concept "classical causality," following Bohr's appeal to "the *classical ideal* of causality," although in the same article and elsewhere, Bohr just uses "causality" to refer to this concept. The reason for adopting the designation "classical causality" is that, as discussed in Section 4, it is possible to introduce alternative, including probabilistic, concepts of causality, applicable in quantum physics, including in relation to complementarity, which Bohr saw as a "generalization of [classical] causality" (Bohr 1987, v. 2, p. 41).

I shall also distinguish classical causality (or other forms of causality considered below) from "determinism," which is an epistemological category, defined by the possibility of predicting the outcomes of such processes *ideally* exactly. Strictly exact measurements and hence the exact verifications of predictions are never possible in practice. In classical mechanics, which deals with individual objects or small classical systems (apart from chaotic ones), both notions in effect coincide. There are, however, theories such as classical statistical mechanics or chaos theory, that are classically causal but not deterministic, in view of the mechanical complexity of the systems considered. This complexity limits us from tracking the behavior of these systems and thus limits us to probabilistic or statistical predictions concerning their behavior.

In the case of quantum phenomena, deterministic predictions are no longer possible, even in considering the most elementary quantum phenomena and events, such as those associated with elementary particles, in contrast to classical statistical physics or chaos theory, where the elementary constituents of systems considered not only behave classically causally but can also be predicted ideally exactly, deterministically. This is because the repetition of identically prepared quantum experiments in all possible cases in general leads to different outcomes, and unlike in classical physics, this difference cannot be diminished beyond the limit defined by Planck's constant, $h$, by improving the capacity of our measuring instruments. This impossibility is manifested in the uncertainty relations, which would remain valid even if we had perfect instruments and which pertain to the data observed in quantum phenomena, rather than to any particular quantum theory. The uncertainty relations are a law defining the character of these data, a law independent of QM, the formalism of which can, however, be rigorously related to the uncertainty relations. Accordingly, as things stand now, the probabilistic or statistical character of quantum predictions must be maintained by realist interpretations of QM or alternative theories (such as Bohmian mechanics).

While, then, in classical physics or relativity, where all systems are classically causal, some of them can be handled ideally exactly, deterministically, and others must be handled probabilistically or statistically, in quantum physics, all systems considered, no matter how elementary, can only be handled probabilistically or statistically. That does not prevent a given theory of quantum phenomena to be, or to be interpreted as, classically causal, in which case what is observed as quantum phenomena, while allowing only for probabilistic or statistical predictions, would be underlain by a classically causal behavior of quantum objects. There are classically causal interpretations of QM, and Bohmian mechanics is expressly classically causal, at the cost of being Einstein-nonlocal. All these interpretations and theories are also, and in the first place, realist. RWR-type interpretations, such as that of Bohr, are not classically causal because the behavior of quantum objects, or the reality thus idealized, cannot be represented or even conceived of, as would be required for classical causality, which implies a law governing it and thus a representation of reality at least in terms of this law.

This primacy of the concept of reality to that of classical causality in RWR-type interpretations, as a concept of reality no longer allowing for classical causality, is apparent in Bohr's passage just cited, by virtue of his claim that "the renunciation of the ideal of causality in atomic physics" is "forced on us" as a consequence of "our not being any longer in a position to speak of the autonomous behavior of a [quantum] physical object." This is a statement concerning the nature of the reality pertaining to, or being idealized by, quantum objects. Bohr's claim requires qualification. In particular, while the situation is different from classical physics, where one can observe such a behavior without appreciably disturbing it, which is no longer possible in considering quantum objects, it is, as noted above, still possible to conjecture or argue for some sort of (even, in principle, classically causal) representation of the independent behavior of quantum objects. One could attempt, and some have, to interpret quantum phenomena and QM in this way, and other theories of quantum phenomena, such as Bohmian mechanics, are both realist and causal. Bohr's statement is careful in that the term "object" is used neutrally as referring to something under investigation, ultimately a form of reality that is beyond representation or conception, an RWR-type reality, and is idealized in terms of quantum objects. The nature of this idealization is subtle, and I shall explain it below. For the moment, as Bohr argued in his reply to EPR, still alongside an appeal to "a final renunciation of the classical ideal of causality," quantum phenomena required "a radical revision of our attitude toward the problem of physical reality" (Bohr 1935, p. 697).

While Bohr, thus, thought that quantum physics required that the classical ideal of causality must be *renounced*, he only thought that our *attitude* toward the problem of physical reality must be *revised*, thus implying that some conception of physical reality may be necessary. In addition, a revision of an *attitude* toward the problem of physical reality is not the same as a revision of a given *concept* of reality, a point to which I shall return in closing this paper. Nevertheless, Bohr did think that quantum phenomena do require such a revision, and he undertook it in his work, adopting the RWR view, as is, again, indicated by his claim that we are not "any longer in a position to speak of the autonomous behavior of a physical object, due to the unavoidable interaction between the object and the measuring instruments." Accordingly, the revision Bohr had in mind, while it required a certain, RWR-type, concept of reality, implied a renunciation of what may be called "the classical ideal of reality," just as it implied, as a consequence, a renunciation of the classical ideal of causality, thus making the quantum-mechanical situation symmetrical in this regard. There are the classical ideal of reality and the classical ideal of causality, and the quantum concept of reality and the quantum concept of causality (defined below), with the corresponding concepts of reality and causality connected in both cases. In the RWR view, the ultimate objects, quantum objects, responsible for quantum phenomena are real, or idealize as what is real, but this reality allows for no physical description or idealized mathematical representation as the objects considered in classical physics or relativity do.

Although quantum *objects* or, again, something in nature so idealized, are thus assumed to exist independently of us and of our measuring instruments, they can never be observed independently, or at least they, say, electrons or photons, have never been observed independently, in contradistinction to quantum *phenomena*, observed in measuring instruments. Nobody has ever observed, at least thus far, an electron or photon as such, in motion or at rest, to the degree that either concept ultimately applies to them, or any quantum objects, no matter how large (and some could be quite large). That is, nobody has observed them insofar as our observational interference by means of measuring instruments, could be neglected and allow us to represent this behavior as independent of this interference in the way it can be done in the case of the objects considered in classical physics or relativity. It is only possible to observe traces, such as spots on photographic plates, left by their interactions with measuring instruments. Such a spot or any such effect can be treated as something *actual*, a representable form of reality, akin to that of classical physics and, again, described by means of classical physics, as against that of quantum objects and behavior, which are equally real but are beyond representation or even conception. Each such trace or a specific configuration of such traces can be treated as a permanent record, which can be discussed, communicated, and so forth.

Accordingly, the recordings of such traces are as objective as they are in classical physics or relativity, except that quantum records can only be predicted probabilistically or statistically, even ideally, while in classical physics, specifically, in classical mechanics, our predictions may be, ideally, exact in some (even a large class of) cases. In quantum physics, only the statistics of multiple identically prepared experiments are repeatable and thus objectively verifiable, which, however, allows quantum physics to be a mathematical-experimental science of nature, just as is classical physics or relativity. It would be difficult or even impossible to do so without being able to reproduce at least the probabilistic or statistical data, which makes it possible, as it is in the case of the experimental data of quantum physics and QM, to objectively verify both these data and themselves and a theory that predicts them.

Although Bohr does not use the language of "reality without realism" (RWR), his view of the structure of quantum measurement as precluding us from being able "to speak of the autonomous behavior of a physical object," is in accord with the concept proposed here under this name. This concept has been introduced by the present author in several earlier works (e.g., Plotnitsky 2015, 2016, 2018a, b, 2019a, b; Plotnitsky and Khrennikov 2015). It is grounded in more general concepts of reality and existence, assumed here to be primitive concepts and not given analytical definitions. These concepts are in accord with most, even if not all (which would be impossible), currently available concepts of reality and existence in realism and nonrealism alike.

By *reality* in general I refer to that which exists or is *assumed* to exist, without making any claim, found in and defining realist theories, concerning the *character* of this existence. The absence of such a claim allows one to place this character beyond representation or knowledge or even conception. I understand existence as a capacity to have effects on the world with which we interact, and the very assumption that something, including the reality of the world, is made on the basis of such effects. Following L. Wittgenstein, I shall understand "the world" as "everything that is the case," thus, the world of events (Wittgenstein 1924, p. 1).

In physics, the primary reality considered is that of matter, keeping in mind that the idea of matter is still a product of thought.[13] "Matter" is, generally, a narrower concept than "nature," although when the ultimate *material* constitution of nature is considered, both concepts merge. Matter is commonly assumed to exist independently of our interaction with it, which implies that it had existed when we did not exist and will continue to exist when we will no longer exist. There are exceptions to this view, including those, such as, most famously, by G. Berkeley, that deny the existence of matter or anything apart from human thought altogether, but these exceptions are rare. The view of matter as existing independently of us is upheld in the RWR-type interpretations of QM, by definition, given the absence of a representation or even conception of this existence. On the other hand, the assumption of the independent existence of matter, while difficult to avoid in physics, is not falsifiable, in contrast to the data observed in measuring instruments or our predictions concerning this data, which (Berkeley had a point there) does not depend on this assumption.

To ascertain observable effects of physical reality entails a representation of them, but not necessarily of how they come about, which implies that a given theory might assume different levels of reality, some allowing for a representation or at least conceptions and others not. Thus, Bohr's interpretation or, following it, the present interpretation of quantum phenomena and QM, or quantum field theory (QFT), assumes that the behavior of the macroworld, as the world of human experience, and specifically of the observable parts of measuring instruments can be represented (in the latter case as treated by classical physics), while quantum objects and behavior, or again, something in nature so idealized, cannot be represented or even allow us to form a conception of them. I qualify by "the world of human experience," because there are macroscopic quantum objects. They cannot, however, *be experienced* as quantum, but can only be established to be quantum by means of measuring instruments, observed classically.

---

[13] A reality, including a reality without realism, can be mental, for example, in mathematics, as discussed in (Plotnitsky 2019c, pp. 203-210).

Realist or ontological thinking and practice in physics, or elsewhere, is defined by the corresponding (realist) theories, which are commonly representational in character.[14] Such theories aim to represent the reality they consider, usually by mathematical models based on *idealizing* this reality by considering those of their features that can be so mathematized and disregarding those that cannot. It is also possible to have a strictly mathematical representation of this reality, but it is difficult to do so without some mediation by physical concepts. I shall discuss this possibility below. All modern, post-Galilean, physical theories are mathematized idealizations, as is quantum theory, in which case, however, it is an idealization that, in the RWR view, does not involve an idealized representation or even conception of the ultimate reality considered. A physical theory may allow one to only have some knowledge concerning the reality it considers without completely describing or representing it. Any such knowledge, however, would require at least a partial representation of this reality.

It is also possible to assume, beyond merely the existence of the ultimate reality considered, that this reality has an independent *architecture* of some sort (possibly temporal in nature), while admitting that it is not possible to either adequately represent this architecture by means of a physical theory or a mathematical model, or even to form a concept of this architecture (beyond that of this reality has some form of architecture), either at a given moment in history or perhaps ever. In the first eventuality, a theory that is merely predictive could be accepted for lack of a realist alternative, but under the assumption that a future theory will do better, in particular by being a properly representational theory based on a workable concept of this architecture and the corresponding representational mathematical model. Einstein adopted this attitude toward QM, which he expected to be eventually replaced by a representational realist theory, specifically a field theory of the type general relativity was. Most of those who make such assumptions conceive of this architecture on the model of the conceptual architecture of classical physics or relativity, while leaving the determination of the specific form of this architecture to the future.

What, then, grounds realism most fundamentally is the assumption that the ultimate constitution of reality possesses properties and the relationships between them, or, as in so-called structural realism (Ladyman 2016), at least a structure emerging from such relationships, that may be either (a) known in one degree or another and *ideally* represented by a given theory or (b) unknown or even unknowable, but still conceivable and, hopefully, eventually to be represented by a corresponding realist theory.[15]

Physical theories prior to quantum theory have all been realist theories. Thus, classical mechanics (used in dealing with elemental individual objects and small classical systems, apart from those considered by chaos theory), classical statistical mechanics (used in dealing, statistically, with large classical systems), or chaos theory (used in dealing with classical systems that exhibit a highly nonlinear behavior) are realist, as concerns the ultimate reality they consider.

---

[14] Although terms "realist" and "ontological" do sometimes designate more diverging concepts, they are generally close in their meaning and will be used, as adjectives, interchangeably for the purposes of the present argument. I shall adopt "realism," as a noun, as a more general term and refer by an "ontology," more specifically, to a given representation or conception of the reality considered by a given realist (or ontological) theory.

[15] One could in principle see the claim concerning merely the existence or reality of something, to which a theory can relate even without representing it, as a form of realism. This use of the term is found or at least suggested in advocating interpretations of quantum mechanics that are nonrealist in the present sense (e.g., Cabello 2017, Fuchs et al 2014, Werner 2014), although, to my knowledge, none of these authors entertained the idea that the ultimate nature of reality considered may (rather than only beyond representation) be beyond conception, possibly ever, which is the strongest form of the RWR view. In any event, I would argue that the present definition of realism or ontology is more in accord with most understandings of either term in the philosophy of physics or elsewhere.

While classical statistical mechanics does not represent the overall behavior of the systems considered because their great mechanical complexity prevents such a representation, it assumes that the individual constituents of these systems are represented by classical mechanics. In chaos theory, which, too, deals with systems consisting of large numbers of atoms, the situation is even more straightforwardly realist, because one assumes a mathematical representation of the behavior of these systems. (The so-called quantum chaos is different, because it is a quantum theory.) The status of these theories as realist could be questioned, on Kantian lines, even in the case of classical mechanics, where the representational idealizations used are more in accord with our phenomenal experience, which, however, as I. Kant argued, does not necessarily correspond to how things, as things-in-themselves, actually are in nature (Kant 1997). Our phenomenal experience can only serve us partially in the case of relativity. This is because, while we can give the relativistic behavior of photons a concept expressible in language and represent it mathematically, we have no phenomenal means of visualizing this behavior, or the behavior represented by Einstein's velocity-addition formula for collinear motion $s = \frac{v+u}{1+(vu/c)^2}$. Nevertheless, all these cases, including relativity, allow for viable idealized realist and causal theories, because we can observe the phenomena considered without disturbing them appreciably by measurement. As a result, we can identify them with the corresponding objects, thus amenable, in their independent behavior, to a representational treatment for all practical purposes, even if the reality itself is ultimately different.

The representation of individual quantum objects and behavior became partial in the so-called old quantum theory (also referred to as semiclassical), in particular Bohr's atomic theory, introduced in 1913 and developed by him and others during the following decade (Bohr 1913). The theory only provided a representation, in terms of orbits, for the so-called stationary states of electrons in atoms, but not for the discrete transitions, "quantum jumps," between stationary states. It was the latter, however, that was Bohr's most decisive and radical step that shapes the development of quantum theory from this point on (Plotnitsky 2019a). The idea became central for Heisenberg, who built on it by abandoning a representation of stationary states as well, which led him to his discovery of QM (Heisenberg 1925). In his initial 1925 assessment of Heisenberg's discovery of QM, by then developed into a full-fledged matrix mechanics by M. Born and P. Jordan (Born and Jordan 1925), Bohr said:

> In contrast to ordinary mechanics, *the new quantum mechanics does not deal with a space–time description of the motion of atomic particles*. It operates with manifolds of quantities [matrices] which replace the harmonic oscillating components of the motion and symbolize the possibilities of transitions between stationary states in conformity with the correspondence principle [which requires that quantum and classical predictions coincide in the classical limit]. These quantities satisfy certain relations which take the place of the mechanical equations of motion and the quantization rules [of the old quantum theory]. (Bohr 1987, v. 1, p. 48; emphasis added)

This assessment, as was Heisenberg's own thinking at the time, was thus based in an RWR-type view and, implicitly, on a corresponding interpretation of quantum phenomena and QM, such as the one expressly offered by Bohr, in terms of complementarity, in "The Quantum of Action and the Description of Nature" in 1929, by that time with the uncertainty relations, introduced by Heisenberg, in place as well (Bohr1987, v. 1, pp. 92-101). I cite this important 1929 article, because, as indicated earlier, in contrast to this article and Bohr's 1925 view just cited, the first (Como) version of Bohr's interpretation, offered in 1927, attempted, still ambivalently, to restore realism and causality to QM. This attempt was quickly abandoned by Bohr, following his discussion with Einstein in October of 1927, which initiated Bohr's path toward his RWR-type interpretation, as is clear in his 1929 "The Quantum of Action and the Description of Nature" (e.g., Plotnitsky 2009, pp. 179-238; 2012, pp. 41-70).[16]

According to Heisenberg himself (back in 1925): "What I really like in this scheme [QM] is that one can really reduce *all interactions* between atoms and the external world ... to transition probabilities" (W. Heisenberg, Letter to Kronig, 5 June 1925; cited in Mehra and Rechenberg 2001, v. 2, p. 242). By speaking of the "*interactions* between atoms and the external world," this statement suggests that QM, as he saw it, was about (predicting) these interactions, observed, as effects, in the measuring instruments involved, a view manifested in Heisenberg's paper and adopted by Bohr in all versions of his interpretation. To return to Bohr's later formulation cited above, we are no "longer in a position to speak of the autonomous behavior of a [quantum] physical object, due to the unavoidable interaction between the object and the measuring instruments which [interaction] in principle cannot be taken into account" (Bohr 1937, p. 87). All that one could say about quantum objects and behavior, or again, the reality thus idealized, could only concern their effects on measuring instruments, probabilistically or statistically predictable by QM, which amounts to the RWR view and a corresponding interpretation of quantum phenomena and QM.

RWR-type interpretations only assume the concept of *reality*, defined as that which is assumed to exist, while, in contrast to realist theories, suspending or even precluding any claims concerning the *character* of this existence beyond representation or knowledge or even conception. At the same time, however, there is no assumption of a uniform or bulk nature of this reality, which is assumed in the RWR view to be different each time, in each experiment (thus, always singular, unique), even though this reality is each time unknowable or even unthinkable, and idealized by the corresponding quantum object at this point of time. As I said, the behavior of the macroworld, as the world of human experience, and specifically of the observable parts of measuring instruments form the other stratum of reality assumed by these interpretations, a stratum that is representable and, in the case of the observable parts of measuring instruments, is treated by classical physics. The existence of the first, RWR-type, stratum of reality is inferred from the totality of effects they have on the world we observe, specifically on measuring instruments. RWR-type reality, idealized in terms of quantum objects and behavior, is both beyond representation or conception and yet is, in each experiment, different. As indicated earlier, the RWR conception of reality as that which is beyond thought is still a product of thought, which makes the impossibility of forming a conception of this reality still a human idealization: ultimately, while it is assumed to exist, to be real, this reality is inconceivable even as inconceivable. But then, so is any other concept of reality, which, however, need not mean, contrary to Berkeley's view, that no material reality exists independently, apart from us. It only means that this existence, however conceived of, is an assumption, a nonfalsifiable assumption, that can, however, be practically justified.

What is, however, the difference between placing the ultimate nature of the reality considered beyond representation or knowledge and placing it beyond conception, beyond thought altogether? For, if, as Bohr says, we are "not being any longer in a position to speak of the autonomous behavior of a physical object, due to the unavoidable interaction between the object and the measuring instrument," would not this imply that this behavior, or again, the RWR-type of reality so idealized, is beyond conception as well? For, if we had such a conception, would we not be able to says something about it? The answer is not straightforward. First, there is difference between some conception or saying something of this reality and forming a more rigorous concept of this reality that would enable us to provide a complete (idealized) representation of it by means of the corresponding theory and the mathematical model used by it. In EPR's phrasing, defining the completeness of a physical theory and representing the idea of a realist theory, so cherished by Einstein, "every element of the physical reality [would] have a counterpart in the physical theory" (Einstein et al 1935, p. 138). I shall call this idea and ideal of completeness "Einstein-completeness." Bohr, by the time of this statement (in 1937) was well aware of this criterion, which he considered in detail in his reply to EPR in 1935, and clearly had it in mind here. While his RWR-type claim would imply that this criterion cannot apply to how the independent, "autonomous," behavior of quantum objects is handled by QM, Bohr saw QM as complete in a different sense: it was as complete a theory capable of predicting quantum phenomena as nature allows such a theory to be, as things stand now. I shall term this concept of completeness "Bohr-completeness." In any event, Bohr clearly makes a stronger claim here: we are no longer in a position to *speak* of the autonomous behavior of quantum objects or the reality thus idealized *at all*, and thus to have a conception of this reality.

---

[16] Both P. Dirac in (Dirac 1958) and J. von Neumann in (von Neumann 1932), followed Bohr's Como argument or, in any event, adopted the same type of epistemology, allowing for realism and causality applicable quantum objects and their behavior, when considered independently of observation, with probabilities only introduced by acts of measurement (Plotnitsky 2009, pp. 197-214; Plotnitsky and Khrennikov 2015, p. 1279). Accordingly, while their views can be seen as in *the spirit of Copenhagen* in other respects, they are not RWR-type views. Published around the same time (Dirac's book was originally published in 1930), these were, arguably, the two most influential books in the history of quantum theory, including (in this case helped by Bohr's Como lecture) as concerns this view, sometimes even seen as "*the* Copenhagen interpretation" (Plotnitsky and Khrennikov 2015, p. 1279). This was not, however, Bohr's interpretation in any of its post-Como versions.

The question then becomes that of whether our inability to do so only (A) characterizes the quantum-mechanical situation as things stand now, or (B) reflects the possibility that this reality is beyond the reach of our thought altogether, ever. While Bohr, thus, at least assumes (A) and while there are intimations that he entertained (B), he never stated so or even stated the possibility of this stronger form of RWR view, which leaves whether he assumed (B) or only assumed (A) to an interpretation. Logically, once (A) is the case, then (B) is possible too. I am not saying that (B) is necessary, but only that it is *interpretively possible*. There does not appear to be any experimental data compelling one to prefer either (A) or (B). Both views are in effect equivalent as far as physics is concerned. They are, however, different philosophically in defining how far our mind can, in principle, reach in investigating the ultimate constitution of nature. (B) puts more radical limits on this reach.[17]

On the other hand, the qualification "as things stand now" applies to (B) as well, even though it might appear otherwise given that this view precludes any conception of the ultimate reality not only now but also ever. It applies because a return to realism is possible, either on experimental or theoretical grounds. This return may take place either because quantum theory, as currently constituted, is replaced by an alternative theory that allows for a realist interpretation, or because (B), or for that matter (A), becomes obsolete even for those who hold it and is replaced by a more realist view with quantum theory in its present form. It is also possible, however, that the RWR view, either of (A) or (B) type, will remain viable in grounding our *interpretations* of QM or QFT, as long as these interpretations remain logically consistent and in accord with the experimental data of quantum physics and not in conflict with experimentally confirmed findings elsewhere. I, again, claim no more for RWR-type interpretations considered here, such as the one adopted here or that of Bohr, who never claimed more either. He also (rightly) argued that Einstein did not in fact challenge this type of interpretation or QM itself on that score (e.g., Bohr 1987, v. 2, p. 57). As I noted, it is also conceivable that a physical theory would emerge, perhaps the one bringing gravity and other forces of nature into a harmony, even if not to unify them, that will require a view that is neither realist nor that of RWR-type.

---

[17] Of course, there is yet another *practical* alternative, that of merely disregarding the question of how quantum phenomena or our predictions concerning them are possible or why it may not be possible to answer this question, or the question of the nature of quantum nonlocality, in short, the question of interpretation of quantum phenomena and QM. This type of attitude was captured by N. D. Mermin's maxim "shut up and calculate," an imperative not adopted by Mermin himself, who also said, on the same occasion: "But I will not shut up," and who made important contributions to foundational questions, including those of entanglement and quantum nonlocality (Mermin 2016, p. 24). In any event, as I am primarily concerned with the question of interpretation, I put this alternative aside.

The nature of the idealization of the ultimate constitution of physical reality in the RWR view of quantum phenomena and QM is very different from that used in classical mechanics, say, in terms of dimensionless massive points mathematically idealizing the motion of material objects in classical mechanics. Elementary particles, such as photons and electrons, are often seen as dimensionless, point-like entities. Indeed, if they had volume, electrons or other charged particles would be torn apart by the electromagnetic force within them). They cannot, however, be considered as point particles in the sense of the idealization of classical mechanics, and when they are understood in terms of *quantum fields*, this concept, too, is very different from that of classical or relativistic fields, such as an electromagnetic field.[18]

The idealization of the ultimate nature of reality in terms of quantum objects in RWR-type interpretations of quantum phenomena is subtle because, while what is *observable* in, or the observable parts themselves of, measuring instruments is always uniquely *classically* defined, what can be considered as the object under investigation or what is considered as a measuring instrument (beyond its observable stratum) in a given experiment is not uniquely defined. According to Bohr:

> This necessity of discriminating in each experimental arrangement between those parts of the physical system considered which are to be treated as measuring instruments and those which constitute the objects under investigation may indeed be said to form a *principal distinction between classical and quantum-mechanical description of physical phenomena*. It is true that the place within each measuring procedure where this discrimination is made is in both cases largely a matter of convenience. While, however, in classical physics the distinction between object and measuring agencies does not entail any difference in the character of the description of the phenomena concerned, its fundamental importance in quantum theory … has its root in the indispensable use of classical concepts in the interpretation of all proper measurements, even though the classical theories do not suffice in accounting for the new types of regularities with which we are concerned in atomic physics. In accordance with this situation there can be no question of any unambiguous interpretation of the symbols of quantum mechanics other than that embodied in the well-known rules which allow to predict the results to be obtained by a given experimental arrangement described in a totally classical way, and which have found their general expression in the transformation theorems, already referred to [Bohr 1935, pp. 697-697n.] (Bohr 1935, p. 701)

It is worth noting first that, although Bohr routinely uses the term "quantum objects" in his writings, here "the objects" is used neutrally referring to something that is different from measuring instruments. The key point is that the difference between measuring instruments, beginning with our bodies and the objects under investigation, which, while it technically exists in classical physics, cannot, unlike in classical physics, be disregarded. Bohr's statement may suggest that, while observable parts of measuring instruments are described by means of classical physics, the independent behavior of quantum objects is described or represented by means of the quantum-mechanical formalism. As noted earlier, however, while this type of view was adopted by Bohr in 1927 and then others, beginning with Dirac (Dirac 1958) and von Neumann (von Neumann 1932), it was not Bohr's view after he revised his Como argument. Bohr does say that observable parts of measuring instruments are physically described by means of classical physics, which nevertheless cannot predict them, again, with a crucial qualification that this description only concerns these observable parts, on which I shall further comment presently. However, he does not say, and does not mean (there is no evidence to conclude otherwise), that the independent behavior of the objects under investigation is described by quantum-mechanical formalism, the "symbols" of which are assumed here, as elsewhere in Bohr, to have only a probabilistically or statistically predictive role. His statement only implies that the objects under investigations in QM cannot be treated physically classically.

---
[18] For discussion of the concept of quantum field from the RWR perspective, see (Plotnitsky 2019a).

This situation is sometimes referred to as the arbitrariness of the "cut" or, because the term [*Schnit*] was favored by Heisenberg and von Neumann, the "Heisenberg-von-Neumann cut." As Bohr noted, however, while "it is true that the place within each measuring procedure where this discrimination [between the object and the measuring instrument] is made is … largely a matter of convenience," it is true only largely, but not completely. This is because "in each experimental arrangement and measuring procedure we have only a free choice of this place within a region where the quantum-mechanical description of the process concerned is effectively equivalent with the classical description" (Bohr 1935, p. 701). In other words, the ultimate (RWR-type) constitution of the physical reality responsible for quantum phenomena observed in measuring instruments is always, in any possible experiment, on the other side of the cut. So are, as part of this reality, those quantum strata of the measuring instruments through which the latter interact with this reality. It is this reality, the reality that is always on the other side of the cut, that quantum objects and their behavior idealize at least in the present interpretation and, I would argue, Bohr's interpretation, at least by the time he adopts an RWR-type view, as he did in his reply to EPR cited here.[19] What is a quantum object, and thus, this idealization, in a given experiment can, as just explained, be different in each case, including possibly something that, if considered by itself, would be classical, as in the case of Carbon 60 fullerene molecules, which were observed as classical and as quantum objects (Arndt et al. 1999). But the RWR-type reality that quantum objects idealize is always on the other side of the cut, and what is responsible for their quantum behavior is defined by their microscopic constitution that can never be on the (classical) measurement side of the cut.

By virtue of their classical nature, then, the individual effects of the interaction between quantum objects and measuring instruments observed in quantum experiments can be isolated materially and phenomenally—we can perceive and analyze them as such, once the experiment is performed. By contrast, at the other end, the ultimate constitution of reality, idealized in terms of quantum objects and behavior, can never be isolated, either physically, insofar as we cannot observe this reality independently, or epistemologically, insofar as we cannot represent or even conceive of this reality. Bohr refers to this situation as the wholeness or indivisibility of quantum phenomena (Bohr 1987, pp. 51, 72-73). This indivisibility precludes us from extracting the independent behavior of quantum objects or the (RWR-type) reality they idealize, thus making us no "longer in a position to speak of the autonomous behavior of a physical object" as opposed to phenomena observed in any quantum experiment.[20]

---

[19] The concept "quantum object" could be defined otherwise, as they would be in a different interpretation of QM or an alternative theory of quantum phenomena, for example, on more realist lines (e.g., Jaeger 2013). (The book considers some nonrealist views as well.)

[20] A somewhat similar argument concerning the stratified character of the reality defining quantum phenomena, if without adopting the RWR view, was proposed in (Rovelli 1996).

Bohr's insistence on the indispensability of classical physical concepts in considering the measuring instruments is often misunderstood, primarily due to insufficient attention to the architecture of Bohr's concepts, as just discussed, in particular by disregarding or missing that, as just explained, the structure of measuring instruments contains both classical and quantum strata. Even though what is observed as phenomena in quantum experiments is beyond the capacity of classical physics to account for them, the classical description can and, in order for us to be able to give an account of what happens in quantum experiments, must apply to the observable parts of measuring instruments. Hence, in the passage in his reply to EPR cited above, he speaks of "the indispensable use of classical concepts in the interpretation of all proper measurements, even though the classical theories do not suffice in accounting for the new types of regularities with which we are concerned in atomic physics" (Bohr 1935, p. 701). He nuances this point later as follows: "[W]e must recognize above all that, even when the phenomena transcend the scope of classical physical theories, the account of the experimental arrangements and the recording of observations must be given in plain language, suitably supplemented by technical physical terminology" (Bohr 1987, v. 2, p. 72). The instruments, however, also have quantum strata, through which they interact with quantum objects, leading to effects that "transcend the scope of classical physical theories." This interaction, which, as discussed in Section 3, is a form of entanglement, is quantum and thus cannot be observed or described as such. It is, in Bohr's words, "irreversibly amplified" to the macroscopic, classical level, say, a spot left on a silver screen (e.g., Bohr 1987, v. 2, p. 73). Bohr, it is true, does not speak in terms of the quantum stratum of the apparatus, but the role of this stratum may be seen as a consequence of what he is saying about the interaction between the objects and the apparatus and the irreversible amplification of this interaction. For, how could an apparatus interact with a quantum object otherwise? In fact, as will be explained in Section 3 as well, what is so amplified in each case is not this interaction but the state of the quantum stratum of the apparatus *after* this interaction had taken place and the object under investigation is elsewhere.[21]

Now, as indicated earlier, the RWR view makes the absence of classical causality nearly automatic. This absence is stricly automatic if one places the ultimate nature of reality beyond conception altogether, because the assumption that this nature is classically causal would imply at least a partial conception of this reality. However, even if one adopts a weaker form of the RWR view, which only precludes a proper (and properly mathematized) representation of this reality, classical causality is still difficult to maintain in considering quantum phenomena. This is because to do so would require a degree of representation (analogous to that found in classical physics) and a law defining the causal relationships that appear to be prevented, in particular, by the uncertainty relations, which are independent of QM. Schrödinger expressed this difficulty in his cat-paradox paper: "if a classical state does not exist at any moment, it can hardly change causally" (Schrödinger 1935a, p. 154).

It is worth considering the reasons for Schrödinger's claim, even at the risk of repeating some well-known features of quantum phenomena, in order to see more clearly how these features appear in RWR-type interpretations of quantum phenomena, although the key experimental facts in question are ascertainable independently. "A classical state" is understood by Schrödinger as defined by the (ideally) exact position and momentum of an object at any moment of time, both of which is no longer possible to ascertain, and hence also to predict, simultaneously in quantum experiments, due to the uncertainty relations. In RWR-type interpretations of the uncertainty relations, beginning with that of Bohr, one not only cannot measure both variables simultaneously but also cannot define them simultaneously for the same quantum object. Thus, the uncertainty relations (however interpreted) prevent us from reconstituting, even if not necessarily from assuming, as one would in an RWR-type interpretation, the trajectory of a quantum object. A given measurement, say, of the position of an object, allows QM to predicts the probability of the outcome of a future measurement of the position of this object at any given future moment of time, but only the probability that this object with be found in a given *area* at this future moment of time.

---

[21] The physical nature of this "amplification" is a separate matter and is part of the (unsolved) problem of the transition from the quantum to the classical, which and related subjects, such as "decoherence," are beyond my scope here.

Technically, anything actually observed can only be observed in a measuring instruments, as a quantum phenomenon, such as, in the case of a position measurement, a spot on a photographic plate, a spot assumed to be a trace of the collision between the object and the screen. As I said, the very existence of quantum objects is inferred from such effects and only from them, although this was not initially realized, because quantum objects, such as electrons or photons, were initially assumed to be classical objects. They still can be treated as classical in certain circumstances, say, when the electron considered is far from the nucleus of the atom. (Their behavior itself still remains quantum and can have quantum effects, not observable in the case of strictly classical objects.) This inference is, moreover, statistical, because in a given single trial of such an experiment, neither the initial emission of the quantum object, defining the first stage of the experiment, nor the fact that the trace thus observed, at the final state of the experiment, is left by the emitted object can be guaranteed. Repeating the experiment with the same initial preparation will lead to a different outcome of the measurement performed after the same time interval. If we consider the possible predictions for future measurement for any set of time intervals following the same initial preparation of the *objects* considered, we can only obtain in Schrödinger's apt language a "catalog of expectation" (Schrödinger 1935, p. 154). I emphasize *objects* (plural) because it is never possible to realize the corresponding experiment for the same quantum object, in the way it is sometimes possible in classical mechanics.[22] Correlatively to the uncertainty relations (which are statistical in character), no such catalog ever allows one to reconstitute the trajectory for a single quantum object in the way this is possible by repeating the same experiment (for either the same or identically prepared objects) in classical mechanics.

Classically, one can continue to perform measurements of both the position and the momentum of an object at any point along its continuous and classical causal trajectory, as a permissible idealization. This is never possible in quantum measurements, even if one assumes, as in Bohmian mechanics, that such a trajectory and, with it, classical causality are possible for a quantum object.[23] In quantum experiments, as against the classical ones, according to Bohr, "a subsequent measurement to a certain degree deprives the information given by a previous measurement of its significance for predicting the future course of the phenomena. Obviously, these facts not only set a limit to the extent of the information obtainable by measurements, but they also set a limit to the meaning which we may attribute to such information" (Bohr 1987, v. 1, p. 18). Thus, whether one deals with the same object or identically prepared different objects, one might speak of how the data obtained or (in M. Scully's quantum eraser experiment [Scully and Drühl 1982) even potentially obtainable in one measurement is made obsolete by another measurement and thus made no longer useful for the purposes of our predictions concerning the subsequent outcome of the experiment. This, returning to Bohr's article cited at the outset of this section, is still due to "the unavoidable interaction between the object and the measuring instrument which in principle cannot be taken into account, if these instruments according to their purpose shall allow the unambiguous use of the concepts necessary for the description of experience" (Bohr 1937, p. 87).

---

[22] R. Feynman's path integral version of QM represents this situation as well, rather than implying that *actual* trajectories, paths, of quantum objects are assumed.

[23] It should be noted that there are arguments for the possibility of classical causality and, in the first place, realism in the case of discrete events (e.g., Sorkin 1991; Smolin 2018, pp. 257-261). In my view, these arguments pose significant problems, beginning with that of explaining the physical mechanism by means of which classical causality can be established in a discrete set. It is not clear that it is possible to conceive of such a mechanism apart from assuming that connections between events are continuous, an assumption that grounds classical causality in classical physics, relativity, and elsewhere, but that was challenged by quantum theory, especially with Bohr's concept of "quantum jumps" (by means of which the electrons would change their energy level in the atoms) in his 1913 atomic theory. This problem is circumvented in the RWR view, by precluding a representation of how quantum phenomena come about and how such predictions are possible, which, admittedly, cannot satisfy those who want realism.

The view itself has governed Bohr's argumentation, including as concerns the point under discussion at the moment, from the introduction of QM on. Thus, as he said, already in the Como lecture:

> It must not be forgotten . . . that in the classical theories any succeeding observation permits a prediction of future events with ever increasing accuracy, because it improves our knowledge of the initial state of the system. According to the quantum theory, just the impossibility of neglecting the interaction with the agency of measurement means that every observation introduces a new *uncontrollable* element. (Bohr 1987, v.1, p. 68; emphasis added)

In his reply to EPR, Bohr speaks of "the finite [quantum] and uncontrollable interaction between the object and the measuring instruments in the field of quantum theory" (Bohr 1935, pp. 697, 700). Accordingly, we can no longer use observation and measurement in the way we do in classical physics to help our quantum predictions. Heisenberg made the same point in his uncertainty relations paper and elsewhere (Heisenberg 1927, pp. 66, 72–76; Heisenberg 1930, p. 36). This point was equally crucial to Schrödinger in his cat-paradox paper (Schrödinger 1935, pp. 152, 154, 157–158).

W. Pauli instructively comments on this situation in his letter to Born (March 31, 1954). The letter refers to Bohr and essentially follows Bohr's argumentation just cited. Pauli first considers a paradigmatic case of classical mechanics, "the determination of the path of a planet":

> . . . if one is in possession of the simple laws for the motion of the body (for example, Newton's law of gravitation), one is able to calculate the path (also position and velocity at any given time) of the body with as high an accuracy as one likes (and also to test the assumed law again at different times). Repeated measurements of the position with limited accuracy can therefore successfully replace one measurement of the position with high accuracy. The assumption of the relatively simple law of force like that of Newton (and not some irregular zig-zag motion or other on a small scale) then appears as an idealization which is permissible in the sense of classical mechanics. (Born 2005, p. 219)

By contrast, in quantum mechanics:

> . . . the repetition of positional measurement in sequence with the same accuracy . . . is of no use at all in predicting subsequent positional measurements. For [given the uncertainty relations] every positional measurement to [the same] accuracy at [a given] time implies the inaccuracy [defined by the uncertainty relations] at a later time, and destroys the possibility of using all previous positional measurements within these limits of error! (If I am not mistaken, Bohr discussed this example with me many years ago.) (Born 2005, p. 219)

Indeed, Bohr's comments cited above present the same line of thought. It follows that our predictions are irreducibly probabilistic, even in the case of individual quantum events, dealing with elementary quantum objects. Nor, by the same token, is it possible to repeat identically prepared experiments with the same outcome, because one experiment does not tell us exactly what will happen in the next experiment, which is identically prepared as concerns the state of the measuring instruments involved (which is possible because these experiments can be treated classically). Suppose we "track" an electron in a hydrogen atom, that is, register the outcome of the corresponding measurements in some measuring apparatus. With Schrödinger's equation and Born's rule in hand, a given measurement allows us to form certain expectations, an expectation-catalog, concerning the future behavior of the electron. However, if we perform a subsequent measurement, the previous measurement becomes meaningless for any future prediction after the second measurement was made. Only the last measurement performed is meaningful for these purposes, and it defines a new expectation-catalog. Correlatively, if we prepare an atom in a certain state in order to make such predictions ("tracking" an electron, again, only amounts to such a preparation), any previous preparation—either in a different state or in the same prepared state (a repeated experiment)—offers no further help in predicting the measurement outcome.

It follows then, whatever assumption one makes concerning the ultimate behavior of quantum objects in interpreting quantum phenomena, that there is an essential difference between classical and quantum phenomena. This difference is defined by "the unavoidable interaction between the object and the measuring instruments which in principle cannot be taken into account," and it can, correlatively, be understood in terms of the impossibility of repeating the same experiment with the same outcome, central for classical physics, from Descartes and Galileo on. One can still repeat the same (individual) experiment in quantum physics, but one cannot in general expect the same outcome. On the other hand, the statistics of multiple trials of the same experiment can be repeated, which allows one to objectively verify both the data thus obtained and a theory, such as QM, and thus to establish QM as a mathematical-experimental science of nature, just as classical physics or relativity is.

Suppose that one performs an experiment on a classical physical object—say, as Galileo did—by dropping a stone from a certain height. One can, at least ideally, repeat the same experiment on the same object or, for all practical purposes, a physically identical object and obtain the same outcome. Such a repetition is always, in principle, possible insofar as one retains a proper record of the experiment. By the same token, the distinction between the behavior of the objects under consideration and the observed phenomena, while present, can be disregarded. In classical physics, this possibility of repeating a given experiment can be destroyed only by literally erasing the data in question, obliterating it without a trace, akin to the burning of the library of Alexandria. As long as the data are still available, the experiment could, in principle, be reconstituted exactly on a given (for all practical purposes identical) object, and in practice this of course happens all the time. In the case of quantum phenomena, one encounters the effects accompanying this type of erasure of the preceding history in any given experiment. While the data necessary to repeat the experiment on an object, such as a photon or electron, are identical to that used in the previous experiment (since all photons or all electrons are indistinguishable from each other), it is, again, in general impossible to repeat any given experiment with the same outcome. Once a measurement is made for the purposes of a prediction, the experiment is closed and the corresponding quantum object is no longer available for these purposes. Even if it is not destroyed physically, it is as good as destroyed for the purposes of any future predictions compatible with this measurement. Conversely, any subsequent measurement establishes a new set of possible predictions, a new expectation-catalog. Accordingly, while a new measurement may not physically destroy the information previously obtained (by "burning" the library of previous expectation-catalogs), it makes this information meaningless for the purposes of further predictions, forming further expectation catalogs, which are defined only by the last measurement performed. Hence, in quantum physics, "a subsequent measurement to a certain degree deprives the information given by a previous measurement of its significance for predicting the future course of the phenomena," as Bohr said (Bohr 1987, v. 1, p. 18).

The set of circumstance just outlined implies a different reason (from those found in classical physics) for our recourse to probability in quantum physics vis-à-vis its use in classical physics. A coin toss is, arguably, the most common example of this difference, even though predicting a quantum event is sometimes compared to a coin toss or (physically, it is essentially the same) a throw of dice. Unless the quantum aspects of the constitution of the coin are considered a factor (which is, generally, not the case), the recourse to probability in this case is necessitated by the mechanical complexities of the situations, because the behavior of a coin, while it cannot be tracked in practice, is composed from and can, in principle, be decomposed into the processes each of which is described by classical mechanics and can be assumed causal, and in the first place, amenable to a realist representation by means of classical mechanics. Accordingly, the outcome is assumed to be predetermined, and given a sufficiently powerful technology, experimental and computational, is in principle predictable exactly, even though it is not and may never be possible in practice. It is true that in the case an individual coin toss (or other cases), we may not have an actual mathematical model, which, while probabilistic, is based on the underlying classical causality (in the way we do in classical statistical physics or chaos theory). Such a model is, however, assumed to be in principle possible, if we, again, disregard the quantum aspect of the situation. Accordingly, even though we cannot track or strictly mathematically model a coin toss, our predictions are not physically spooky, because, unlike in the case of quantum phenomena, in an RWR-type interpretation, we assume a causal process connecting each toss and its outcome. By the same token, it is, in principle, possible to construct a representational mathematical model (perhaps a digital one) of a coin's behavior, similar to models we have in chaos theory, even though it may never happen in practice.

This is never possible in the case of quantum objects, no matter how elementary, such as elementary particles, because no such tracking is even in principle possible, because we cannot disregard in the interference of measurement. Nor, by the same token, is any decomposition of a quantum behavior into parts *observationally* possible (even if one assumes a realist or causal interpretation of QM itself or another theory predicting quantum phenomena, such as Bohmian mechanics). This is because any attempt to do so will always involve the same situation of measurement and prediction, and thus a new individual quantum phenomenon, and not a part of the previous phenomenon. Any attempt to "cut through" a phenomenon can only produce yet another closed individual phenomenon, leaving quantum objects themselves inaccessible inside phenomena. As Bohr said: "In fact, the *individuality* of the typical quantum effects finds its proper expression in the circumstance that any attempt of subdividing the phenomena will demand a change in the experimental arrangement introducing new possibilities of interaction between objects and measuring instruments which in principle cannot be controlled" (Bohr 1987, v. 2, p. 40). Hence, each quantum event or phenomenon is irreducibly individual, unique in itself and discrete in relation to any other quantum event or phenomenon, and cannot be decomposed into a sequence of simpler subevents responsible for it, in contrast to physically classical random events, such as a coin toss. Physically, the latter is a sequence of many events and is, in principle, a subject of a realist causal classical description. It is only the mechanical complexity of the underlying situation defining a given event that prevents us from predicting the outcome of the event exactly.

That, again, does not in itself mean that an underlying realist or causal interpretation of QM or a realist and causal theory of quantum phenomena is not possible. In this case, one would of course also avoid the spooky character predictions at a distance, spooky because there is no theory of how they physically come about. As things stand now, however, no such interpretation or theory can avoid the experimental situation just described either. This situation would for example obtain (along with the uncertainty relations or the statistical nature of quantum predictions) in Bohmian mechanics, albeit, again, at the cost of Einstein-nonlocality. Given the nature of quantum phenomena, as just described, a realist and causal interpretation or theory can only concern what *underlies* quantum phenomena, because there is no way, thus far, to track the behavior of quantum objects themselves, in the way we in principle can in considering individual objects in classical physics. This possibility leads to the assumptions underlying the use of probability and statistics in classical physics, even if we have no (mathematized) physical theory of the behavior of classical objects which we predict probabilistically, as in the case of a coin toss. No such independent tracking is even possible in quantum physics because we cannot separate the behavior of quantum objects, no matter how elementary, from their interactions with measuring instruments and only observe the effects of these interactions. Hence, as noted, while in classical physics or relativity, where all systems are classically causal, some of them can be handled deterministically and others must be handled probabilistically or statistically, in quantum physics all systems considered, no matter how elementary, can only be handled probabilistically or statistically. Nor do they need to be assumed to behave causally, and they are not in RWR-type interpretations. As Bohr argued, essentially in view of the considerations just given:

> [I]t is most important to realize that the recourse to probability laws [in quantum physics] is essentially different in aim from the familiar application of statistical considerations as practical means of accounting for the properties of mechanical systems of great structural complexity. In fact, in quantum physics we are presented not with intricacies of this kind, but with the inability of the classical frame of concepts to comprise the peculiar feature[s] of the elementary processes. (Bohr 1987, v.2, p. 34)

Referring to "the peculiar feature of the *elementary processes*" has additional significance because, in contrast to classical mechanics, exact predictions are no longer possible, even ideally or in principle, even in considering the behavior of the elementary quantum objects, "elementary particles." That is, exact predictions concerning the effects of their interactions with measuring instruments are no longer possible.

"The classical frame of concepts" may appear to refer to the concepts of classical physics, and this frame does include these concepts. However, by this time (in 1949), Bohr adopts the RWR view, which places the ultimate nature of reality responsible for quantum phenomena beyond conception, at least as things stand now, and possibly (although it is, again, not clear whether Bohr has been willing to go that far) beyond the reach of thought altogether. As explained, the qualification "as things stand now" still applies but in this case only insofar as this form itself of the RWR view may be abandoned by a future theory of quantum phenomena. This position gives the phrase "the classical frame of concepts" a broader meaning: all concepts that we can form (ordinary, physical, or philosophical) are classical. The question is only whether our concepts could one day become applicable in quantum theory or whatever can replace it.

Purely mathematical concepts are a possible exception. Realizing this, Heisenberg, eventually moved from a form of RWR view he adopted at the time of his discovery of QM to a form of mathematical realism in his later thinking. He continued to maintain certain affinities with Bohr's views, in particular, by assuming that QM does not offer a physical description of quantum objects and behavior in space and time and that one cannot speak about them by using ordinary language and concepts (e.g., Heisenberg 1962, pp. 145, 178-179).[24] Bohr, by contrast, equally rejected the possibility of a mathematical representation of quantum objects and behavior, or, again, the reality they idealize, along with a physical one, in all versions of his interpretations, apart from a brief attempt, mentioned earlier, at a more realist approach in the Como lecture. It is true that Bohr often speaks of quantum objects and behavior as being beyond our phenomenal, representational intuition, also involving visualization, sometimes used, including by Bohr, to translate the German word for intuition, *Anschaulichkeit* (e.g., Bohr 1987, v. 1 p. 51, 98-100, 108; v. 2, p. 59). This impossibility of visualization or (which is another phrase often used by Bohr) pictorial visualization of quantum objects and their behavior is important, also in its juxtaposition of the idea of "*Bild*" or mental image of Hertz and Boltzmann, discussed in the Appendix. It is clear, however, that, apart from the Como lecture, Bohr saw the ultimate nature of the reality responsible for quantum phenomena as being beyond any representation and thus conception, including a mathematical one, or any possible analysis, at least as things stand now. As he said in his arguably most definitive statement on the subject: "In quantum mechanics, we are not dealing with an arbitrary renunciation of a more detailed analysis of atomic phenomena, but with a recognition that such an analysis is *in principle* excluded" [beyond a certain point] (Bohr 1987, v. 2, p. 62).

Bohr's argumentation leading to this conclusion, outlined in the same article and discussed here, was, as Bohr noted (with much regret), outright rejected, rather than contested by Einstein: "Even if [Einstein's realist] attitude might seem well balanced in itself, it nevertheless implies a rejection of the whole argumentation exposed in the preceding," an argumentation "aiming to show" Bohr's RWR-type conclusion just cited, culminating in this strong phrase "*in principle* excluded," with Bohr's emphasis on "*in principle*" (Bohr 1987, v. 2, p. 62).

By the time he reaches the ultimate version of his interpretation, Bohr defines atomic or quantum phenomena, strictly in terms *effects*, observed, in classically described parts of measuring instruments, as a result of their interaction with quantum objects or, again, the reality thus idealized. As he said on the same occasion:

> I advocated the application of the word phenomenon exclusively to refer to the *observations* obtained under specified circumstances, including an account of the whole experimental arrangement. In such terminology, the observational problem is free of any special intricacy since, in actual experiments, all observations are expressed by unambiguous statements referring, for instance, to the registration of the point at which an electron arrives at a photographic plate. Moreover, speaking in such a way is just suited to emphasize that the appropriate physical interpretation of the symbolic quantum-mechanical formalism amounts only to predictions, of determinate or statistical character, pertaining to individual phenomena appearing under conditions defined by classical physical concepts [describing the relevant observable parts of measuring instruments]. (Bohr 1987, v. 2, p. 64)

Referring, phenomenologically, to "observations," rather than, ontologically, to the observed situations themselves, explains Bohr's (Kantian) choice of the term "phenomenon," although it may still be better, in order to avoid any confusion, to speak of "quantum phenomena," thus defined. The key features of Bohr's concept are clearly apparent here, in particular, the fact that the term refers "to the observations obtained under specified circumstances," and hence, only to already registered phenomena. Because the observed parts of the measuring arrangement are described classically, phenomena and measuring instruments, as classical objects (as against, quantum objects, responsible for phenomena), could be considered as identical, as they always are in classical physics. By contrast, the *emergence* of these phenomena, which is due to the interaction between measuring instruments and *quantum objects*, or quantum objects and behavior themselves, or, again, the reality they idealize, are no longer available to a representation by means of QM or otherwise, at least as things stand now.

---

[24] See (Jaeger 2019), for an assessment of Heisenberg's overall later views of QM, including in (Heisenberg 1962).

The epistemological cost of the RWR view is, again, not easily absorbed by many physicists and philosophers, and to some, beginning, famously, with Einstein, is outright unacceptable. This is not surprising because the features of quantum phenomena manifested in many famous experiments, such as, paradigmatically, the double-slit experiment, and those that led to RWR-views defy our basic assumptions concerning the workings of nature and thought. These assumptions, arising due to the neurological constitution of our brain, have served us for as long as human life itself, and within certain limits are unavoidable, including in physics, although, while respected by classical physics, they were already challenged by relativity. QM and then QFT have made this challenge much greater, albeit without entirely renouncing these features, which, in the RWR view, are applicable at levels other than that of the ultimate constitution of nature, specifically in describing by classical physics the behavior of measuring instruments.

Thus, it is natural and even humanly unavoidable to assume that *something happens* between observations, given changes that we observed in the physical states of the instruments used. Indeed, the sense that something happened is one of the most essential elements of human thought. However, in the RWR-type view, the assumption that "something happened" is at most provisional and ultimately inapplicable in considering the independent behavior of quantum objects, or again, the reality thus idealized. Anything that can actually *happen*, any quantum *event*, can only be encountered as manifested in the observable parts of measuring instruments, at which level classical concepts would apply. According to Heisenberg:

> There is no description of what *happens* to the system between the initial observation and the next measurement. …The demand to "describe what happens" in the quantum-theoretical process between two successive observations is a contradiction in adjecto, since the word "describe" [or "represent"] refers to the use of classical concepts, while these concepts cannot be applied in the space between the observations; they can only be applied at the points of observation. (Heisenberg 1962, pp. 47, 145; emphasis added)

The same would apply to the word "happen" or "system," or any word we use, whatever concept it may designate, including, again, reality. Bohr is reported to have said: "We must never forget that 'reality' too is a human word" (Kalckar 1967, p. 234). Although one should always be careful as concerns reported statements, especially by Bohr, this statement is in accord with his argumentation in his writings. While, as Bohr often noted, quantum physics cannot avoid using ordinary language, most especially in describing quantum experiments, it imposes new radical limits on our use of it. Our language and concepts cannot apply to quantum objects and behavior, or again, the reality they idealize, the reality of the ultimate constitution of nature, as understood by quantum theory in the RWR view. Heisenberg acknowledges these limits in the same book: "But the problems of language are really serious. We wish to speak in some way about the structure of the atoms and not only about 'facts'—the latter being, for instance, the black spots on a photographic plate or the water droplets in a cloud chamber. *But we cannot speak about the atoms in ordinary language*" (Heisenberg 1962, pp. 178-179; emphasis added). Nor is it possible in terms of ordinary concepts, from which ordinary language is indissociable, or even in terms of physical concepts, assuming that they can ever be free from ordinary language and concepts.

This is a formidable problem even if one adopts the RWR view. The term "reality" does not pose a difficulty in this regard, because it has no concept associated with it and hence does not belong to ordinary language and concepts, or even to physical concepts. In a way, it functions almost as a mathematical symbol. It is true that, as noted earlier, this view itself is not falsifiable, but then neither is any realist ontology of "the ultimate constitution of reality," which is only a human expression as well. A greater difficulty is the expression "quantum objects *interact* with each other" between experiments, for example, leading to an entanglement. As will be discussed in detail in the next section, in considering a measurement I shall speak of the interaction and entanglement between the quantum object and the quantum stratum of the instrument, which interaction takes place before anything is actually observed, as an effect of this interaction. Thus, this interaction, too, occurs before an observation.

The way to handle this difficulty in the RWR view, still in provisional terms, is as follows. Although one makes an assumption, a nonfalsifiable assumption, of some form of relation between two or more quantum objects (which may include quantum strata of measuring instruments), there is no classical concept, such as "interaction" or even "relation," applicable to this "relation." There are only certain observable effects that are describable or numerically representable (some of which may be correlated) and that could be predicted by using the mathematical formalism of QM, specifically those parts of the formalism that we associate with both this assumption and the concept of entanglement (entangled state vectors). These predictions are falsifiable. How these effects come about remains beyond representation, including by such terms as "interaction" or "relations."

As indicated, the assumption that one cannot speak about the structure of atoms in ordinary language or even in terms of physical concepts, still allows for the mathematical representation of what happens between the experiments because mathematics is, as Heisenberg said on an earlier occasion, "fortunately" free from the limitations of ordinary language and concepts, fortunately because one could take advantage of this freedom, as Heisenberg did in creating QM, on which he commented at the time (Heisenberg 1930, p. 11). At the time of his discovery of QM, Heisenberg, adopting a form of the RWR view, used this freedom to invent a theory, QM, designed to only predict the probabilities of events observed in quantum experiments. By contrast, in his later writings Heisenberg assumed the possibility of such a mathematical representation of the ultimate constitution of reality, while, thus giving his view a Platonist bent, excluding any other language or concepts, ordinary or even physical, as possibly applicable to this constitution (Heisenberg 1962, pp. 145, 167-186). This possibility also reveals the importance of the qualification that the reality considered is only *idealized* in terms of quantum objects and behavior, because if this representation is strictly mathematical such physical concepts as behavior or even object, or quantum, may not apply. Heisenberg speaks of this representation in terms of symmetry groups and defines "elementary particles" accordingly, which is to say, without considering them as "particles" in any physical sense, which is a form of the RWR view. On the other hand, the concept of elementary can be given a mathematical sense insofar as the corresponding representation of the symmetry group considered is irreducible (Wigner 1939). For Bohr, a mathematical representation of this reality is assumed to be no more possible that a physical one, at least as things stand now, an assumption adopted here as well.

 Thus, in RWR-type interpretations, QM is incomplete by Einstein's criterion of completeness, "every element of the physical reality must have a counterpart in the physical theory" (Einstein et al 1935, p. 138), because it offers no representation of the ultimate constituents of nature and their individual behavior. I shall call theories that do so "Einstein-complete," in parallel with the Einstein-local, and Einstein wanted fundamental theories to be both (the combination referred to as local realism), and ideally, he wanted them to be classically causal or even deterministic. As indicated earlier, however, QM was seen by Bohr as complete in a different sense, which I termed earlier "Bohr-complete." It was as complete a theory capable of predicting (nonrelativistic) quantum phenomena as nature allows a theory of these phenomena to be, as things stand now. There is no change in this respect. QM remains our standard theory of these phenomena, and while it is quantum-nonlocal in allowing for statistical correlations between distant quantum events, it is, or at least may be interpreted as, Einstein-local.

### 3. What does a quantum measurement measure and what does a quantum theory predict?

According to Bohr, then, "the unavoidable interaction between the object and the measuring instruments which in principle cannot be taken into account" precludes us from being able to separate the behavior of quantum objects from this interaction and to represent this behavior independently in the way it is possible to do in classical physics or relativity (Bohr 1935, p. 697).

This situation has momentous implications for the nature or structure of quantum measurement and predictions concerning quantum phenomena. These predictions become unavoidably probabilistic or statistical, but, as explained in the preceding section, for reasons entirely different from those that dictate the recourse to probability and statistics in classical physics. As Bohr eventually came to realize (although he did not use the term entanglement) and as I shall now explain, in any given experiment, the quantum object under investigation and the measuring instrument used become entangled. I say this keeping in mind the qualifications made earlier concerning the provisional nature of using, in RWR-type interpretations, the term "interaction" or "entanglement" as referring to the independent behavior of quantum objects, including the quantum strata of measuring instruments. There is no conception that we can form that would be rigorously applicable to this "interaction" or this "entanglement," which, as indicated earlier, is technically between the quantum stratum of the measuring instrument and the quantum object under investigation. We can only rigorously define "entanglement" in terms of the mathematical formalism of QM and make the corresponding predictions by using this formalism, supplemented by Born's rule, which, while (because it essentially amounts to using the complex conjugation) it is reasonably natural given the mathematical formalism of QM, is nevertheless added to this formalism, rather than being inherent in it. Why it works is enigmatic. Why Born's rule works may even be the greatest mystery of QM. We know *how* it works, but not *why*, at least as things stand now.

With these qualifications in mind, the entanglement, in each measurement, between the quantum object considered and the measuring instrument used defines quantum measurement and enables us, by using QM, or some other theory, to predict probabilistically or statistically (no other predictions are, again, possible on experimental grounds, as things stand now) the outcomes of certain other possible experiments, observed strictly in measuring instruments. Which outcomes can be predicted depends on what measurement we perform, say, that of momentum, which allows us to predict (probabilistically or statistically) the outcomes of future momentum measurements, but, in view of the uncertainty relations or complementarity, irrevocably precludes making the position measurement at the same moment of time and, thus, making any *predictions* concerning future position measurements. We *can make* a position measurement instead of a momentum one at a later point in time, but its outcome *cannot be predicted*, if the momentum measurement was made. One needs to make a position measurement at the initial stage of the experiment, thus changing its setup, in order to do so.

It follows that any *possible* future phenomenon, which is actualized when and only when the corresponding experiment is performed (otherwise it is not a phenomenon in Bohr's definition) is thus always discrete in relation to the one defined by the initial measurement enabling prediction of this phenomenon, without, in Bohr's or other RWR-type interpretations, assuming the possibility of represent or even conceive of a physical process connecting phenomena. No such process can be assumed given that we are no "longer in a position to speak of the autonomous behavior of the physical object" between observations, which also precludes classical causality. By the same token, our quantum predictions are predictions at a distance, are quantum-nonlocal. On the other hand, there is no need to assume an action at a distance, which allows us to see these predictions as Einstein-local, although they are, again, sometimes interpreted in terms of Einstein-nonlocality.

In making a measurement, say, again, a position measurement in quantum physics, one always deals with the entanglement of the quantum object and the measuring instrument, physically the quantum stratum of the measuring instrument (as the observable parts of the instrument are treated classically), as *two quantum systems*, in contrast to an *interaction of two classical systems*, the object and the instrument, which defines classical measurement. In the classical case, we can disregard or properly control the interference of a measuring instrument, and treat a given classical object as a single independent system. We can also simultaneously establish the position of the object, because the interference of the measuring device enabling this position measurement in not affected by the measurement of the object's momentum, and could be disregarded as well. In the case of a classical momentum measurement, we thus measure the momentum of the object itself or at least may, ideally and in principle, assume that we do, while, as I shall explain presently, in quantum measurement we in fact measure the momentum of the instrument after its interaction with the object, the interaction that entangles them. By the same token, in classical mechanics (in considering individual or small systems, apart from chaotic ones), we can predict both the momentum and the position ideally exactly, deterministically at any future point of time and can reconstitute the value of both in the past.

This is never possible in the case of quantum phenomena, in view of the uncertainty relations, even if one assumes that both variables can actually be assigned to quantum objects themselves, as in the case of Bohmian mechanics. By the same token, QM only gives us information, in general probabilistic, about the future, and never about the past, as each new measurement makes all preceding measurements meaningless as concern our predictions after this new measurement has taken place. Any available past information is only that obtained by measurements. As Schrödinger said, in considering quantum measurement in terms of entanglement in his cat-paradox paper, "the rejection of realism … imposes obligations" (Schrödinger 1935, p. 158). While Schrödinger, disparagingly, saw QM, at least in an RWR-type of interpretation, as "a doctrine born of distress" (Schrödinger 1935, p. 154), he offered important insights into this "doctrine" in many of its aspects: nonrealism (even though he stopped short of reaching a full-fledged RWR view), discreteness, quantum nonlocality, and most especially entanglement, the concept he introduced there in both German (*Verschrankung*) and English. He also addressed complementarity, even if without using the term. As he said:

> From the standpoint of the classical model the momentary statement content of the $\psi$-function is far from a complete description. From the new standpoint it must be complete [as complete as possible]. It must be impossible to add to it additional correct statements, without otherwise changing it. … Thence it follows that two different catalogs, that apply to the same system under different conditions or at different times, may well partially overlap but never so that one is entirely contained within the other. For otherwise it would be susceptible to completion through additional correct statements, namely through those by which the other one exceeds the first.—The mathematical structure of the theory automatically satisfy this condition. There is no $\psi$-function that furnishes exactly the same statements as another and in addition several more statements. (Schrödinger 1935, p. 159; translation slightly modified)

"Complete" refers here to Bohr-completeness: QM is as complete a theory of predicting quantum phenomena, of providing expectation-catalogs for them, as nature allows us to have in accordance with all experimental data available, at least as things stand now. It is also consistent with other theories, such as relativity included, implying Einstein-locality. On the other hand, QM is, as Schrödinger, also indicate here, it is incomplete from "the standpoint of the classical model," in other words, is Einstein-incomplete. If one adopts the RWR view, the interaction between the object and the measuring instrument, and the resulting momentum measurement only helps us to have an expectation-catalog provided by the formalism, say, $\psi$-function (cum Born's rule), for possible future momentum measurements. This catalog is exhaustive: it cannot be supplemented by a catalog for future position measurements, as reflected in the uncertainty relations and complementarity, both implicit in Schrödinger's statement. In some cases, such as those of the EPR-type, the corresponding predictions could even be made with probability one, and moreover, on basis of a measurement performed not on the object for which the prediction is made but another object, with which that object previously interacted. Still, in contrast to classical physics, these predictions can only concern either one or the other complementary variables, but never both together in the same experiment. In addition, as explained in detail below, they do not guarantee that the value thus predicted, with probability one, corresponds to the reality that will obtain in the measurement thus predicted unless this measurement is actually performed. This is because one can always perform a complementary measurement at that future moment in time. They are still only *predictions* and not statements defining what is real, or what is guaranteed to be real, as such predictions may, in principle, be in classical physics.

The nature of an interaction between the object and the measuring apparatus (technically, again, the corresponding quantum stratum of the apparatus) as an entanglement contains important further features. This interaction itself is not measurement: it occurs before the measurement takes place, even if usually, but not always (because it can be delayed), quickly following this interaction. Once performed, the measurement, say, that of the momentum, disentangles the object and the instrument, with the observed outcome "irreversibly amplified" to the level of *the classically observed stratum of the apparatus* (Bohr 1987, v. 2, p. 73). This outcome is actually associated with the quantum stratum of the apparatus after the interaction with the object rather than with the object itself, which is a crucial point, discussed below. It is this disentangling that, like in any other case of entanglement, enables us to predict, by using QM, the probability that the next momentum measurement at a given point of time will be within a certain range (Schrödinger 1935, pp. 162-163). Alternatively, if the initial measurement was that of the position, one could predict the probability that a future position measurement will locate the position within a certain area. This future measurement will create a new entanglement, which will enable future predictions, but, as explained earlier, the expectation-catalog for them will be different and will deprive the previous expectation-catalog of any value for predictions after this new measurement (Schrödinger 1935, pp. 159-160).

Observed quantum phenomena, which also disentangle the entanglements between objects and instruments, are never entangled. Only quantum objects and $\psi$-functions that we use to predict their future interactions with measurement instruments are. But we can never observe either: quantum objects because they can never be physically observed in themselves, and $\psi$-functions because they never represent either quantum objects apart from measurement or measurements themselves, at least in the RWR view. Nor are phenomena, or quantum objects, can ever be in a linear superposition (a key feature of the formalism); only state vectors can, making a linear superposition a purely mathematical feature of QM, as opposed to a classical wave superposition, from which the term is borrowed. At most one could speak, still not altogether rigorously, of the superposition of *possible* measurement outcomes or phenomena, probabilistically predicted by using $\psi$-functions (cum Born's rule). Indeed, $\psi$-functions never represent measurements either, even if one takes a realist view of them as representing what happens between measurements. A measurement makes a $\psi$-function "collapse" or, in Schrödinger's more precise language, which also better fits the RWR view, makes "[to be] no more" (Schrödinger 1935, p. 161). The language of collapse suggests and usually represents a realist position, according to which the $\psi$-function is at least part of the physical representation of the independent behavior of quantum objects. On the other hand, in the RWR view, because a $\psi$-function is only part of the mathematical machinery enabling us to have our expectation-catalogs, once the measurement associated with the expectation-catalog defined by a previous measurement and the $\psi$-function used for creating this catalog is performed, this $\psi$-function is no longer of any use for future predictions. "It is no more!" A new $\psi$-function is necessary for the catalog defined by this new measurement.

The special circumstances of the EPR experiment allow us to make such predictions with probability one, but the complementary nature of the predictions concerning the variables in question, such as the momentum and the position (or the opposite direction of the spin in the Bohm-Bell version of the experiment for spin), will still be in place. Both are never possible within the same experiment, which fact, while underappreciated by EPR, grounded Bohr's argument in his reply to EPR, on both counts, the completeness and the Einstein-locality of QM, although the predictions involved are predictions at a distance and thus are quantum nonlocal. As I shall explain below, however, one can *ideally* reconstitute the EPR case from the standard measurement case, by making simple additional measurements, only ideally because the special ideal state used in the EPR experiment for continuous variables is not normalizable, which prevents the corresponding experiment from being actually performed.

In contrast to the EPR-type situation (where, as a result, we can make predictions with probability one), in general we don't know the entangled state, say, again, defined by the $\psi$-function, of the composite system consisting of the object and the quantum stratum of measuring instrument involved. This is because the measurement immediately disentangles this state, thus enabling us to compile the corresponding expectation-catalog concerning future possible experiments for the desired variable, while, again, depriving us of doing so for a complementary variable. Any prediction in this catalog is quantum nonlocal, a prediction at a distance, because at the time of measurement the object in question is elsewhere, and thus any prediction concerning it is a prediction at a distance, without, as in the classical case, having a physical description and the corresponding mathematical representation of the process connecting these two events. A $\psi$-function (cum Born's rule) may provide a catalog for predictions at any point on the temporal continuum, but it only provides a catalog for possible measurements, and not a continuous mathematical representation of the process that leads to a future measurement. All quantum phenomena are, thus, discrete in relation to each other, a discreteness that implies quantum nonlocality, as predictions at a distance, without implying an action at a distance. Quantum nonlocality could, again, be interpreted in this Einstein-nonlocal way, but need not be.

The deeper essence of this situation is revealed by the fact, noted by Bohr, that, because the measurement as such physically takes place after the interactions between the object and the apparatus has already occurred, *the alternative, complementary, measurement* is always possible following *the same interaction* between and entangling the object and the instrument. In other words, this interaction and this entanglement does not by itself define the measurement (which disentangles the entanglement) that one can perform. Instead, it allows for either of the two complementarity measurements to be performed *after this interaction*, although both measurements can never be performed together in the same experimental arrangement. These two facts define the complementarity of these two measurements and the resulting phenomena. Each measurement requires a different experimental arrangement, which, it follows, can be made *after* this interaction has already take place. According to Bohr:

> After a preliminary measurement of the momentum of the diaphragm [with slits through which a particle can pass, as in the double-slit experiment], we are in principle offered the choice, when an electron or photon has passed through the slit, either to repeat the momentum measurement or to control the position of the diaphragm and, thus, to make predictions pertaining to alternative subsequent observations. It may also be added that it obviously makes no difference, as regards observable effects obtainable by a definitive experimental arrangement, whether our plans of considering or handling the instruments are fixed beforehand or whether we prefer to postpone the completion of our planning until a later moment when the particle is already on its way from one instrument to another. (Bohr 1987, v. 2, p. 57)

The first point confirms that each type of measurement or the corresponding prediction requires a separate specified arrangement, which grounds most of Bohr's responses to Einstein's criticism of QM, including in the case of EPR's argument. It might be added, given that the point is not made by Bohr himself, that this requirement also implies that one needs two separate quantum objects to realize both measurements and the corresponding predictions, again, central to Bohr's countering EPR's argument. Bohr's second, subtler, point, which is under discussion at the moment, is that it is always possible "to postpone the completion of our planning until a later moment when the particle is already on its way from one instrument to another." This observation also anticipates the possibility of Wheeler's delayed choice experiment, in fact presented by Wheeler via Bohr's interpretation of quantum phenomena and QM in terms of his concept of phenomenon, an experiment amplifies the role of quantum nonlocality (Wheeler 1983, pp. 182-192).

If, however, one always makes a measurement after the object has left the location of the measurement after interacting with the instrument and is on the way to another possible measurement, what does one, then, measure, given that the object itself is no longer here to be measured? One measures, *the state of the measuring instrument*, either its momentum or its position, as both, again, can never be measured in the same arrangement, in view of the uncertainty relations.[25] More accurately, either measurement reflects the state of the quantum stratum of the instrument, which interacted with the object in the past (however recently, but always in the past), the state amplified to the classical level of the observation, to which and only to which even such concepts as "momentum" or "position," and hence the uncertainty relations, can rigorously apply physically in the RWR view of the situation. Nor, indeed to begin with, do we ever observe quantum objects as such, a fact stressed throughout this paper. However, given that one can perform either measurement after the object has left the location of the apparatus, one can ascertain, regardless of an interpretation, both key features in question: (a) that one can perform either measurement concerning the state of the quantum stratum of the apparatus, with the outcome amplified to the classical level of the observable part of this instrument; and correlatively (b) a quantum-nonlocal nature of any possible prediction, based in either measurement thus performed, concerning any future measurement. One only makes a prediction concerning a distant future event to which one is not physically connected at the time of the measurement enabling this prediction, because this measurement only pertains to the quantum stratum of the apparatus, while the object is already on its way to a distant, spatially separated, area where a measurement corresponding to the prediction can be made. In order to verify this prediction itself, given its probabilistic or statistical nature, the experiment must be repeated with the same preparation multiple times.

---

[25] This point appears to have been missed or, in any event, not addressed in commentaries on Bohr, or by treatments of quantum measurement elsewhere. Subtle as it is, Schrödinger's analysis of quantum measurement in his cat-paradox paper does not consider this point (Schrödinger 1935, pp. 158-159). Von Neumann's analysis comes close to this point, but, while it is conceivable that von Neumann realized it, he did not expressly make it as such, and indeed, as explained below, some of his statement suggest a different, more realist, view of the situation, whereby the measured quantity is in fact attributed to the object at the time of measurement (von Neumann 1932, pp. 355-356). This point also suggests yet another challenge to the thought experiment known as "Heisenberg's microscope," illustrating the uncertainty relations (Heisenberg 1927). Heisenberg's analysis has been challenged previously on several grounds, but not, to my knowledge, by assuming that a quantum measurement measures the state of the measuring instrument after its interaction with the object and not the state of the object itself.

One might assume (without affecting how the experiments are performed or how formalism is used), say, in the case of the momentum measurement, the exchange of momenta between the object and the instrument, that the momentum of the object will correspond to the difference between two momentum measurements of the instruments and, thus, measure the momentum of the object itself (e.g., von Neumann 1932, p. 355). Physically, however, we never measure that momentum, given, first, that, the object has already left the location of measurement, and secondly, that we could have performed instead the position measurement, a circumstance missed or, in any event, not noted by von Neumann. The same applies to measurements that verify our predictions, which are always recorded corresponding to the state of the measuring instrument, by an amplification from the corresponding quantum stratum of the instrument to the observable classical part of it. This is why, in the RWR view, no physical properties, such as position and momentum, are assumed to be applied to quantum objects or to the reality thus idealized, but only to the observed, classically described, parts of measuring instruments.

Thus, using the outcome of the measurement representing the state of the apparatus, one can predict, quantum-nonlocally, by means of the $\psi$-function (cum Born's rule), a possible outcome of a future measurement on or, rather, again, involving the object, without "in any way disturbing the system," just as we would in the EPR type experiment (Einstein et al 1935, p. 138). It is true that there was an interaction between the object and the instrument before that measurement. But this is also the case for the two objects of the EPR pair, which have been in an interaction, which entangled them, before the last stage of measurement on one or the other object of the pair, which enables us to make a prediction concerning the other. It is important to keep in mind that such quantum-nonlocal predictions of complementarity variables require, in either the standard or the EPR case, two different specified arrangements (and two different objects) to perform both experiments, in accordance with how we decide to set the apparatus. As discussed in Section 5, EPR, do not take these circumstances into account, which allows Bohr to argue that EPR's use of the phrase "without in any way disturbing the system" is ambiguous and ultimately inapplicable in the EPR experiment, at least not without additional qualifications, not offered by EPR.

In the case of the standard, rather than EPR-type, case, our predictions are, again, not with probability one, but the essence of the situation is the same. Indeed, with some simple additional arrangements, following the first measurement, one can reproduce the (idealized) EPR case, which led Bohr to realize the essential parallel between the EPR experiment and the standard quantum measurement (Bohr 1935, pp. 699-700; Bohr 1938, pp. 101-103; Bohr 1987, v. 2, p. 60). The nature of the initial interaction as an entanglement is, however, most crucial here, as Schrödinger realized. Bohr did not use the language of entanglement, but the situation is, nevertheless, that of an entanglement between the object and the measuring instrument, technically, again, the quantum stratum of it through which it interacted with the object. Even in the standard quantum measurement, we are unavoidably dealing not only with the object under investigation, as in classical physics (because there we can disregard the role of measuring instruments), but with a composite entangled quantum system, consisting of the object and the measuring instrument.[26]

To return to Schrödinger's idiom, the unknown "quantum state" (the unknown $\psi$-function) of the combined entangled system, the object and the measuring instrument, resulting from their interaction allows one to make, *after this interaction* and hence when the object is elsewhere, either of the two complementary measurements, determining either of the two complementary quantities. Each would allow one to compile, by means of a $\psi$-function, the corresponding catalog for future possible measurements. This makes it tempting to think that both quantities are "already there" to be measured and a combined catalog encompassing both is to be formed, which would make QM incomplete because it does not allow for such a possibility, as there are, as we have seen, no such expectation-catalogs (Schrödinger 1935, p. 159). *Either is not the same as both*. One can assume that quantum objects *possess both variables*. However, because each measurement requires a separate, differently specified, arrangement, incompatible with the other, there is no experiment, including the EPR type, that would allow one to measure or predict both variable. Thus, there is no way to claim, as EPR do, that one can actually *predict* more than QM can predict, and thus to prove that it does not "[exhaust] the possibilities of observation" and hence is incomplete, Bohr-incomplete (Bohr 1987, v. 2, p. 57). This reasoning governs Bohr's reply to EPR.

The entanglement between the object and the measuring instrument, as any entanglement, allows one to *potentially* have more probabilistic knowledge that an actually performed measurement can, probabilistic because Schrödinger, again, notes in this case "each individual statement or item of knowledge is after all a probabilistic statement" (Schrödinger 1935, p. 160). Epistemologically, an entanglement is always "an entanglement of predictions" and, as "an entanglement of our knowledge" is only that of probabilistic knowledge, even if sometimes these predictions are made with probability one (Schrödinger 1935, p. 160). The determinate knowledge is only given by a measurement, and then there is no longer any entanglement, unless of course it is a future one that happens after a measurement, but such a new entanglement will only be an entanglement of predictions. In the present situation (of the entanglement between the quantum object under investigation and the measuring instrument used), this entanglement allows one to *potentially* establish two expectations-catalogs, because either disentangling measurement can be performed. But it does not ever allow one to ever *actually* have both catalogs and combine them into a single expectation-catalog. The two catalogs are mutually exclusive and, thus, require *two* mutually exclusive arrangements, and two separate quantum objects or, as in the EPR case, pairs, to realize these catalogs.

This potentiality of having either catalog is a subtle illustration of Schrödinger's claim that "the best *possible* knowledge of a whole does not necessarily include the same for its parts," which and thus entanglement itself, he sees as "the great difference over against the classical model theory," in fact, in his view, the defining trait of quantum physics (Schrödinger 1935, p. 161;

---

[26] I am indebted to M. G. D'Ariano for drawing my attention to the significance of always considering composite systems in quantum theory, the point emphasized in his recent works (some of which are cited here), in contrast to other quantum-informational approaches to quantum foundations, which tend to underappreciate this significance.

emphasis added). I shall not discuss this point as such here, except by reiterating that it is crucial that this knowledge is only a *possible* knowledge. This is crucial because it is, again, never possible to have an expectation-catalog that can encompass two alternative expectation-catalogs that the entanglement of the objects and the instruments potentially enables. This potentiality could only be actualized as one of the two possible mutually exclusive events and two mutually exclusive sets or catalogs of predictions concerning future events that may follow each of these events. Understanding quantum measurement in terms of entanglement and the resulting parallel with the EPR case, again, reveals that in any quantum experiment we always consider composite interactive systems, and never a single system.[27]

The situation also amplifies the nonlocal nature of quantum prediction. This is because deciding, after the interaction on one or the other complimentary measurement, made at time $t_1$ enables alternative, mutually exclusive, predictions concerning a distant object, or rather a distant experiment, at time $t_2$, and neither measurement any longer requires interference with the object, because the latter has already left the location of the instrument. In contrast to classical physics, there is no way to ascertain that both variables can ever be assigned, because this assignment would require representing the behavior of the object independently of its interaction with measuring instruments, which is never possible. These circumstances make this interaction, irreducible in quantum physics, correlative to the nonlocal nature of quantum predictions, as spooky predictions at a distance (spooky, again, insofar as we do not know why they work), while maintaining Einstein-locality, and thus avoiding a spooky action at a distance. One can at time $t_2$ perform an alternative complementary measurement on the object. This will, however, disable the possibility of ascertaining the value of the first variable and disable the original prediction. This is part of the conceptual architecture of complementarity, to which I now turn.

## 4. Complementarity and quantum causality

The main lineament of Bohr's concept of complementarity, especially as referring to the mutually exclusive nature of certain quantum phenomena, should be apparent from the preceding discussion. Defined arguably most generally, complementarity is characterized by:
   (a) a mutual exclusivity of certain phenomena, entities, or conceptions; and yet
   (b) the possibility of considering each one of them separately at any given point; and
   (c) the necessity of considering all of them at different moments for a comprehensive account of the totality of phenomena that one must consider in quantum physics.

The concept was never quite presented by Bohr in a single definition of this type. However, the characterization just stated may be surmised from several of Bohr's statements, such as the following one: "Evidence obtained under different experimental conditions cannot be comprehended within a single picture, but must be regarded as *complementary* in the sense that only the totality of the phenomena [some of which are mutually exclusive] exhaust the possible information about the objects" (Bohr 1987, v. 2, p. 40). In the RWR view, this information is, again, comprised of the data observed in measuring instruments, which defines *quantum phenomena* in Bohr's sense, as effects of the interaction between measuring instruments and *quantum objects*, or the reality that quantum objects idealize. This reality is beyond knowledge or even conception and about it we can, thus, have no information of any kind. Parts (b) and (c) of this definition are just as important as part (a), and disregarding them often leads to misunderstandings. The fact that at any given point either of the two complementary measurements could be performed and the corresponding phenomenon established is just as crucial as the mutual exclusivity of the two complementary phenomena considered. That we have a free, or at least a sufficiently free, choice concerning what kind of experiment we want to perform is part of the very idea of experiment in science, including in classical physics (Bohr 1935, p. 699). However, in contrast to classical physics (or relativity), in quantum physics implementing our decision concerning what we want to do enables us to make certain types of possible predictions and will irrevocably exclude certain other, *complementary*, types of predictions, all probabilistic or statistical in character, in accordance with the uncertainty relations, which are statistical in character as well. The uncertainty relations are not the same as complementarity, but they can be given, as they were by Bohr, an interpretation in terms of complementarity. Thus, if one makes a position measurement, one can, probabilistically or, if one repeats the same measurement a new (rather than repeats it immediately, in which case the outcome will be ideally the same) statistically make, form an expectation-catalog for, predictions concerning future position measurements, say, as registered by a spot on the screen. Doing so, however, irrevocably precludes one from making any predictions, or forming an expectation-catalog, concerning future momentum measurements on the same object.

---

[27] This aspect of quantum theory is reflected in the purification postulate introduced in (D'Ariano et al 2017, pp. 168-168), which expressly builds on Schrödinger's argument concerning the fundamentally quantum nature of our knowledge concerning entangled systems: "the best *possible* knowledge of a whole does not necessarily include the same for its parts." See also note 26 above.

Rather than arbitrarily selecting one or other part of physical reality that can be assumed to be already there, as in the case of classical mechanics, our decisions actively shape the single reality observed in the measuring instrument used and what can be predicted, even if not always what can actually happen in the future, and preclude the complementary alternative for either the measurement or for future predictions. I qualify, because, as indicated earlier, if, say, on the basis of the momentum measurement at time $t_1$, one made a prediction concerning a momentum measurement at time $t_2$, which prevents one from making any predictions concerning the position of the objects at $t_2$, one could still perform the position measurement at $t_2$, thus not only depriving one of the possibility of verifying the original prediction but also alternatively defining the reality at $t_2$. While, however, our decision to perform a particular measurement cannot give us certainty concerning what *will* happen, what *may or may not* happen in quantum experiments still depends on our decision which experiment to perform. This is because, unlike in classical physics, we cannot assume that both measurements represent *parts* of the same single reality observed in measuring instruments. The ultimate reality, idealized in terms of quantum objects, responsible for what is observed cannot itself be observed or, in the RWR view, represented or even conceived of.

Speaking in classical terms, one could say that there are only parts which no longer add up to a whole of the type found in classical physics. The very essence of complementarity is, however, that one cannot, at least in the RWR-type of interpretation, think in classical terms, even though we can use the corresponding variables themselves classically in describing the behavior of measuring instruments. We can never apply both of them (position and momentum) at the same time, because the uncertainty relation and complementarity apply to measuring instruments, and only to them. Each measurement creates the only observed reality there is, defined by our decision of which measurement to perform, and the alternative decision would establish the other reality, with the ultimate reality responsible for what is observed manifested only in its effects on measuring instruments, thus, stratifying the overall architecture of the reality considered. By the same token, each measurement defines what we can and, in considering complementary situations, cannot predict concerning future observable reality and thus predict what *might* happen. Again, however, a measurement can allow us to estimate but cannot define *what will happen*, *what will be real*, which can only be defined by another measurement, and one can always perform an alternative measurement (complementary to the one the outcome of which was predicted) at this future point and thus define a different observed reality from the one predicted.

It may be noted that wave-particle complementarity, with which the concept of complementarity is associated most commonly, had not played a significant role in Bohr's thinking, especially after the Como lecture (where one might find statements that can be interpreted in terms of this complementarity). First of all, Bohr was always, even in the Como lecture, acutely aware of the difficulties of applying either concept to quantum objects. Bohr's ultimate solution to the dilemma of whether quantum objects are particles or waves was that they were neither. Instead, either "picture" refers to one of the two complementary sets of discrete individual *effects* of the interactions between quantum objects and measuring instruments, particle-*like* effects, which may be individual or collective, or wave-*like* effects, which are always collective, while composed of discrete individual effects. The example of the latter are "interference" effects, composed of the large number of discrete traces of the collisions between the quantum objects and the photographic screen, in the double-slit experiment in the corresponding setup (when both slits are open and there are no means to allow us to know, even in principle, through which slit each object has passed).

The concept of complementarity is better exemplified by complementarities of spacetime coordination and the application of momentum or energy conservation laws. There are two complementarities here: the first is that of the position and momentum measurements, and the second is that of the time and the energy measurements. These complementarities are correlative to Heisenberg's uncertainty relations and establish Bohr's interpretation of them. Technically, the uncertainty relations, $\Delta q \Delta p \cong h$ (where $q$ is the coordinate, $p$ is the momentum in the corresponding direction), only prohibit the simultaneous *exact* measurement of both variables, which is always possible, at least ideally and in principle, in classical physics, also allowing one to maintain classical causality there. As Bohr said, in his reply to EPR:

> In the phenomena concerned [including those of the EPR type] we are not dealing with an incomplete description characterized by the arbitrary picking out of different elements of physical reality at the cost of [sacrificing] other such elements, but with a rational discrimination between essentially different experimental arrangements and procedures which are suited either for an unambiguous use of the idea of space location, or for a legitimate application of the conservation theorem of momentum. Any remaining appearance of arbitrariness concerns merely our freedom of handling the measuring instruments, characteristic of the very idea of experiment. (Bohr 1935, p. 699)

This situation is, thus, defined by the irreducible role of measuring instruments in the constitution of quantum phenomena, here mutually exclusive ones, in contradistinction to classical physics, where such "elements of reality" as the position and the momentum can always be attributed simultaneously to the object itself under investigation. Bohr borrows the phrase "elements of reality" from EPR's paper, in responding to which this point is made (Einstein et al 1935, p. 138).

By the same token, the uncertainty relations are not a manifestation of the limited accuracy of measuring instruments, because, as noted, they would be valid even if we had perfect instruments. In Bohr's interpretation, the uncertainty relations make each type of measurement involved in them complementary: mutually exclusive yet allowing us a freedom of performing either of them at any moment in time. Furthermore, in Bohr's interpretation, one not only cannot measure both variables simultaneously but also cannot define them simultaneously. According to Bohr: "the statistical character of the uncertainty relations in no way originates from any failure of measurement to discriminate within a certain latitude between classically describable states of the objects, but rather expresses an essential limitation of applicability of classical ideas to the analysis of quantum phenomena" (Bohr 1938, p. 100). As he said elsewhere: "we are of course not concerned with a restriction as to the accuracy of measurement, but with a limitation of the well-defined application of space-time concepts and dynamical conservation laws, entailed by the necessary distinction between measuring instruments and atomic objects" (Bohr 1987, v. 3, p. 5). This limitation is defined by the complementary nature of these two applications. The situation is correlative to the necessary recourse to probability or statistics in considering quantum phenomena, which is, as explained earlier, due to "the inability of the classical frame of concepts to comprise the peculiar feature[s] of the elementary processes" (Bohr 1987, v. 2, p. 34). Although classical concepts do apply to quantum phenomena, this application is limited by the uncertainty relations and complementarity.

Complementarity may, thus, be seen as a reflection of the fact that, in a radical departure from classical physics or relativity, the behavior of quantum objects of the same type, say, electrons, is not governed, individually or collectively, by the same physical law, in all possible contexts, specifically in complementary contexts. Speaking of "*physical* law" in this connection requires caution, because, in Bohr's interpretation, there is no physical law representing this behavior, and even no probabilistic law concerning the outcomes of individual quantum experiments. The behavior of quantum objects leads to mutually incompatible observable physical effects in complementary contexts, which are probabilistic or statistical in nature. On the other hand, the mathematical formalism of QM offers correct probabilistic or statistical predictions (no other predictions are, again, possible) *in all contexts*.[28]

Hence, as a quantum-theoretical concept, complementarity acquires probabilistic or statistical aspects, as well as, ultimately, mathematical aspects (because one needs mathematics to make these predictions), rarely addressed in considering complementarity. Khrennikov's articles (Khrennikov 2019b, c, 2020a) are a welcome exception as concerns probability. Indeed, complementarity may be seen, as it was by Bohr, as a probabilistic or statistical generalization of the idea of causality in the absence of classical causality. Bohr never explained the nature of this generalization, but it can be understood by means of the concept of quantum causality, introduced by this author previously (Plotnitsky 2011, Plotnitsky 2016, pp. 206-206, Plotnitsky 2018a).

I define "quantum causality" as *the probabilistic or statistical determination* of what may happen in a possible future observation (it need not be, technically, a measurement), at time $t_2$ as *a result of what has happened* previously happened as a quantum event, for example, that of the measurement defined by our decision concerning which experiment to perform at a given moment in time, $t_1$. Both emphasized phrases are crucial because, as explained earlier, one can, at $t_2$, perform an alternative measurement and thus, by *another decision*, establish a reality different from the one predicted at $t_1$, even if this prediction was made, as in EPR-type experiments, (ideally) with probability one. Thus, one can, on the basis of the position measurement at $t_1$, make a prediction concerning an outcome of a possible position measurement at $t_2$ and then perform instead a momentum measurement at $t_2$. However, because this alternative outcome could not have been predicted at the time when the initial measurement was made and is not a result of our decision made in performing the initial experiment, the definition of quantum causality just given remains intact. Whatever is registered as a quantum event defines a possible set of, probabilistically or statistically, predictable future events, as outcomes of a certain set of possible future experiments, and by complementarity, precludes certain other types of predictions, although it is, again, not possible to *establish* the reality thus predicted at the time, $t_2$, for which the prediction is made. This can only be done by a measurement at $t_2$, and because an alternative measurement can be performed, a reality different from the predicted one could be brought about at $t_2$. All such predictions are, again, quantum-nonlocal, are predictions at a distance, but they respect Einstein-locality and thus preclude any physical action at a distance.

---

[28] This situation is also responsible for what is known as "contextuality," which is a statistical concept, while complementarity may be defined apart from probability or statistics. However, when used by Bohr in QM, complementarity is a statistical concept. I considered contextuality from the RWR perspective in (Plotnitsky 2019b).

Quantum causality is a fundamentally probabilistic concept. In contrast to classical causality, in which case what has happened determines what *will happen*, in effect connecting all events involved in a single causal chain or network (even though, due to our lack of knowledge of how this happens, our predictions could still be probabilistic), quantum causality determines what *may (or may not) happen*, and *not what will happen*, possibly, and in the RWR view definitively, in the absence of the underlying classically causal connections between events. Classical causality is ontological or realist, while quantum causality is, in RWR-type interpretations, while it arises by virtue of quantum reality as an RWR-type reality, is probabilistic and as such is *epistemological*. It pertains strictly to our interactions with the world by means of our experimental technology and our thinking. It has to do with our knowledge, even though the nature of quantum probability itself is *nonepistemic*, because the recourse to probability is not due to the lack of our knowledge concerning how quantum events ultimately come about. The knowledge pertaining to quantum causality only concerns the data obtained in already performed experiments (or other quantum events, which may be seen as experiments performed by nature) and possible probabilistic predictions concerning future events, with both type of data being manifested strictly in phenomena (in Bohr's sense), either already registered or *possibly* to be registered in the future. In the RWR view, there is, again, no knowledge or even conception of the reality responsible for what is so observed. The recourse to probability is, again, not due to *the lack of knowledge concerning the nature of reality*, a classically causal reality, as in classical statistical physics, but is due to *the ultimate nature of this reality* itself, which is of RWR-type and hence cannot be classically causal. If one reverts to classical mechanics, our predictions concerning individual objects become, ideally, deterministic and, correlatively, there is no complementarity.

This definition of quantum causality is in accord with recent views of causality in quantum information theory (e.g., Brukner 2014, Hardy 2010, D'Ariano 2018), except that it expressly brings in complementarity, rarely considered in these arguments. M. G. D'Ariano, in (D'Ariano 2018), defines causality in physics in general by means of this type of concept, which is also consistent with complementarity, although the latter is not expressly considered by D'Ariano. Classical causality as defined here or determinism (the term used by D'Ariano as essentially equivalent to classical causality) is merely a special case of causality in this probabilistic sense, the case which allows for *ideally* exact predictions in classical mechanics. This is a justifiable view and an important concept, providing a very general definition of causality, applicable beyond quantum physics or even physics, for example, in biology, evolutionary theory, neuroscience, decision science, or economic. Just as the present concept of quantum causality, this concept is manifestly consistent with both Einstein-locality and quantum nonlocality, as defined here.

I adopt the term "quantum causality" rather than speak of "causality" as such, as D'Ariano does, in part for historical reasons, given the previous use of the term causality, including by Bohr, in the sense of classical causality, although it is inviting and effective to see classical causality just as a limit case of probabilistic causality found in quantum theory, or of causality in D'Ariano's sense. In addition, quantum causality as defined here and, to begin with, the renunciation of classical causality (a realist concept) are grounded in the RWR view. While D'Ariano's concept is consistent with the RWR view or allows for an RWR-type interpretation, it does not appear that D'Ariano subscribes to the RWR view in this or related articles, but instead appears to adopt a form of mathematical ontology akin to that of Heisenberg in his later works, as mentioned earlier (e.g., D'Ariano 2017). This is a possible and in many ways attractive position, which also has certain affinities (which is not to say coincides) with structural realism (e.g., Ladyman 2016). By contrast, the RWR view may be seen in terms of *structural nonrealism* (Plotnitsky 2018b). Besides, as indicated in discussing Heisenberg's views, this position, at least, in Heisenberg's and D'Ariano's cases, may still be seen as a form of reality without realism, insofar as there is no physical concept that can capture this ultimate nature of reality, thus suggesting an interpretation of D'Ariano's title "Physics without physics" of (D'Ariano 2017). I am not attributing this interpretation to D'Ariano, but only suggesting it as a possible interpretation.

D'Ariano's concept is expressly linked to the arrow of time. The arrow of time is, however, also part of quantum causality as defined here. It may be added that, although the case is not considered by D'Ariano, performing a measurement of the type alternative to the initial measurement and the corresponding future predicted by it can only happen at a future moment, which preserves the arrow of time at the level of phenomena. In the present, RWR-type, view, the arrow of time is only manifested classically in the observable phenomena. D'Ariano, on the other hand, *appears* to see the arrow of time as found in the ultimate workings of reality responsible for quantum (or classical) phenomena, at least in (D'Ariano 2018), although it is possible that he would accept the present view.

The present view does not, however, imply that at the ultimate level of reality there is no change or multiplicity but only permanence and oneness. This view is sometimes found in literature (e.g., Barbour 1999, Gomes 2016), in part advanced because of the lack of classical causality in dealing with quantum phenomena and the fact that the equations of QM or QFT (or of classical physics or relativity) are mathematically time reversible. In the RWR view, however, this type of representation or this concept would not apply to the ultimate constitution of reality, any more than concepts of change, becoming, motion, or space and time, or indeed any concepts. In the RWR view, the equations of QM or QFT, such as Schrödinger's equation or Dirac's equation, are not equations of motion of quantum objects, but are mathematical structures providing (along with Born's or analogous rules) quantum-nonlocal expectation-catalogs concerning the outcomes of possible quantum experiments, which requires the assumption of the arrow of time. What can, in the RWR view, be *objectively* ascertained is that the ultimate nature of reality is such that all our interactions with it, on all scales, by means of experimental technology (beginning with that of our bodies), including in dealing with quantum phenomena, entail the arrow of time, which quantum causality reflects.

With these considerations in mind one can understand Bohr's view of complementarity as a generalization of causality. On the one hand, "our freedom of handling the measuring instruments, characteristic of the very idea of experiment" in all physics, our "free choice" concerning what kind of experiment we want to perform is essential to complementarity (Bohr 1935, p. 699). On the other hand, as against classical physics or relativity, implementing our decision concerning what we want to do will allow us to make only certain types of predictions and will exclude the possibility of certain other, *complementary*, types of predictions. Complementarity generalizes causality in the absence of classical causality and, in the first place, realism, because it defines which events, which reality of events, are established by a measurement and which events can and cannot then be probabilistically or statistically predicted by our decision concerning which experiment to perform. Complementarity, as a quantum-mechanical concept is, again, defined as such by the irreducible role of measuring instruments in the constitution of quantum phenomena and, hence, the RWR view, cannot be properly understood apart from probability. Conversely, however, the way probability and, with it, quantum causality, works in quantum physics cannot be understood apart from complementarity.

It is instructive to consider how complementarity and quantum causality work in the case of predictions with a "probability equal to unity," which, as will be seen in the next section, becomes crucial to countering EPR's argument, where this phrase plays a key role, as does the phrase "without in any way disturbing the system" (Einstein et al 1935, p. 138). I adopt this phrase as my title in this paper, but only as reflecting the fact that one can make EPR-type *predictions* at a distance (by performing a measurement on another system with the system concerning which such a prediction is made had previously been in interaction), but not because, contrary to EPR's criterion of reality, there are, prior to a measurement, "elements of reality" corresponding to such predictions. Their criterion of physical reality (in effect assumed in all of Einstein's argument concerning the subject) is as follows: "*If, without in any way disturbing a system, we can predict with certainty (i.e., with probability equal to unity) the value of a physical quantity, then there exists an element of physical reality corresponding to this physical quantity*" (Einstein et al 1935, p. 138; EPR's emphasis). According to EPR, this "sufficient" criterion equally applies in classical and quantum physics, a claim challenged by Bohr, who argued that, while this criterion is (ideally) applicable in classical physics, it contains "an essential ambiguity" when applied, even ideally, to quantum phenomena, including those of the EPR type, as EPR tried to do (Bohr 1935, p. 698).

One might ask first: What does it mean to predict with probability equal to unity in physics? It means that we assumed that, if we measure the predicted quantity by means of some measuring instrument, this measurement confirms our prediction. Does this quantity correspond to an element of physical reality, unless the measurement is performed? Technically, this is not so even in classical physics, because there could be factors that will make it no longer possible to obtain the predicted value by a measurement. Rigorously speaking, no predictions with probability equal to unity could guarantee that what is so predicted will happen. In classical physics or relativity, however, this is not a serious objection to EPR's criterion, because it applies ideally in principle, unless there is an outside interference.[29] A prediction with probability equal to unity is an idealization because the value of the predicted quantity cannot be measured exactly or always predicted exactly. This idealization is, however, permissible in classical mechanics in dealing with small systems (apart from chaotic ones), because one can repeat each individual experiment and improve our measurements with each repetition, in principle indefinitely.

The situation is essentially different in the case of quantum phenomena because of the irreducible role of measuring instruments in their constitution and complementarity, which in effect disables Einstein's criterion or at least makes it ambiguous, even in an ideal experiment, such as the EPR experiment. First of all, we cannot repeat quantum experiments in the way we can in classical physics or improve the precision of our measurement beyond the limits establish by the uncertainty relations, which, as I said, would apply even if we had ideal instruments. As will be seen in the next section, this impossibility of repeating a given quantum experiment with the same outcome plays an implicit role in the EPR-type experiments, a role unperceived by EPR. All we can numerically improve is our *calculations of probabilities* of our predictions. For the moment, suppose that one had predicted, on the basis of a position measurement at time $t_1$ a future value of the position of an object at time $t_2$ with probability equal to unity, which is possible, as in EPR-type experiments, by means of a measurement performed on a different quantum object and thus without in any way disturbing the first one. This prediction can then always be confirmed, again, within the limit of idealization, by the corresponding position measurement at $t_2$, in accord with EPR's criterion of reality. However, as explained earlier, measuring at $t_2$ the value of the complementary variable, that of momentum, instead of the predicted one, that of the position, will make it impossible to assign the position variable to the object at $t_2$, even though if we had measured the position instead it would strictly correspond to our prediction. There is no experiment that could allow us to do so, as opposed to classical physics, where we can, in principle, always measure and, correspondingly, predict both variables simultaneously, and correlatively, ideally assign both properties to the object itself considered independently of measurements. In other words, a prediction with probability equal to unity is applicable, even ideally, only if this prediction is in principle verifiable, which cannot be assured in considering quantum phenomena, including those of the EPR type, in the way it could be in principle in classical physics.[30] These are these considerations that ultimately led Bohr to adopt the view that only a measurement, which establishes a registered phenomenon, and never a prediction, even with probability equal to unity, defines what is real at both levels, phenomena and quantum objects or, again (and here this qualification becomes especially important), the RWR-type reality they idealize, a reality that we cannot observe or represent, or even conceive of.

---

[29] One might also note, along Bayesian lines, that predictions with any probability are only meaningful insofar as those who made them or know of them are still alive.

This analysis of predictions with probability equal to unity in quantum physics helps one to argue that the EPR experiment confirms that, although quantum-nonlocal, QM is *both* complete, Bohr-complete (insofar as it predicts anything that is possible to predict, as things stand now) and Einstein-local, rather than shows that QM is *either* Bohr-incomplete *or* Einstein-nonlocal, as EPR contended. This contention, Bohr argued, was open to challenge because EPR and Einstein in his subsequent communications underappreciated the role of measuring instruments and complementarity in defining quantum phenomena.

### 5. Reality, complementarity, and quantum nonlocality in the EPR experiment

During the last half a century, following Bell's and the Kochen-Specker theorem and related findings, the debates concerning quantum foundations has shifted towards quantum correlations and quantum nonlocality, vis-à-vis the completeness of QM, although the questions of completeness and realism have remained an unavoidable background of these debates. Neither EPR's original article nor Bohr's reply used the term "entanglement," introduced by Schrödinger, in responding to EPR's paper, or spoke in terms of correlations. On the other hand, the question of locality and completeness, and the alternative between them, was central to the Bohr-EPR exchange and to Einstein's subsequent communications, compelling him to speak of "a spooky action at a distance" [*spukhafte Fernwirkung*] (e.g., Born 2005, p. 155). Most of the key findings and arguments involved in these debates deal with discrete variables and Bohm's version of the EPR experiment. The main reason is that, unlike Bohm's version, the original thought-experiment proposed by EPR cannot be performed in a laboratory. This does not affect the fundamentals of the case, which can be made by considering the original EPR experiment as an idealized thought experiment. Bohm's version of the EPR experiment can and has been performed, confirming the existence of quantum correlations, which can, thus, be ascertained experimentally, apart from QM or any quantum theory.[31]

---

[30]That a prediction with probability equal to unity is not the same as establishing the reality of what is so predicted has been also stressed by quantum Bayesians (QBists), but on the grounds of the subjective nature of Bayesian probability, rather than the type of reasoning used here (e.g., Fuchs et al 2014; Mermin 2016, pp. 231-238).

Ironically, realist (Einstein-complete) theories that would predict EPR-type correlations appears to be Einstein-nonlocal (or alternatively, violate other assumptions generally regarded as basic) in view of Bell's and the Kochen-Specker theorems, and related findings, which, again, thus far deal with discrete variables. Among the most famous of these findings are those of D. M. Greenberger, M. Horne, A. Zeilinger, and L. Hardy, and, from the experimental side, A. Aspect's experiment and related experimental work, such as that by A. Zeilinger and his group (Aspect et al 1982, Greenberger et al 1989, Greenberger et al 1990, Hardy 1993).[32] As I said, the meaning of these findings, or the EPR experiment and Einstein's and Bohr's arguments concerning it, have been intensely debated. I shall bypass these debates, in part because my argument deals with the subject via complementarity, underappreciated in these debates.[33] I would argue, however, that at stake in all of these findings are situations that are governed by complementarity and that can be considered from an RWR-type perspective. In order to support this argument, I shall reexamine the key features of Bohr's reply to EPR, based on his (RWR-type) interpretation of quantum phenomena and quantum mechanics, as considered in the preceding analysis.

Bohr contested Einstein's argumentation by offering this interpretation and, most especially, by analyzing in detail the roles of measuring instruments and complementarity. It was Bohr's analysis of these roles, which, he argued, were underappreciated by EPR, that allowed him to conclude that QM "would seem to fulfill, within its scope, all rational demands for completeness," at least Bohr-completeness, insofar as it predicted everything that could be predicted in accordance with the experimental evidence available (Bohr 1935, pp. 696, 700n; also Bohr 1987, v. 2, p. 57). He also argued that QM fulfills these demands without sacrificing Einstein-locality, by virtue of the compatibility of Bohr's argument and, thus, his interpretation of quantum phenomena, including those of the EPR type, and QM with "all exigencies of relativity theory," which implies Einstein-locality (Bohr 1935, p. 701n).

---

[31] There are experiments (those involving photon pairs produced in parametric down conversion) that statistically approximate the EPR experiment for continuous variables. I cannot consider these experiments here, but they may be argued to be consistent with the present argument. They also reflect the fact that the EPR experiment is a manifestation of correlated events for identically prepared experiments with EPR pairs on the model of the Bell–Bohm version of the EPR experiment. The analysis of the latter experiment involves further subtleties, including those related idealizations and "degrees" in which one can assess these correlations and, thus, quantum nonlocality, which are related to the strict numerical limits on Bell's and related inequalities. Although there have been many helpful treatments of these subjects since, N. D. Mermin's papers on the subject, assembled in (Mermin 1990, pp. 81-185), remain one of the best sources in capturing these subtleties, and are notable for their clarity and elegance of exposition.

[32] I only cite some of the key earlier experiments. There have been numerous experiments performed since, some in order to find loopholes in these and related subsequent experiments.

[33] The literature dealing with these subjects is nearly as immense as that on interpretations of QM. Among the standard treatments are (Bell 2004, Cushing and McMullin 1989, Ellis and Amati 2000). N. D. Mermin, again, offers a particularly lucid treatment (Mermin 1998, pp. 81-185). See also (Brunner et al 2014), for a current assessment of Bell's theorem. As noted, these theorems and most of these findings pertain to quantum data as such, and do not depend on QM. It should also be kept in mind that there are realist and causal views of quantum entanglement and correlations, either in realist interpretations of QM, more conventional (e.g., Kupczynski 2018), or more esoteric, such as the many worlds interpretation, in alternative theories, such as Bohmian mechanics, which is Einstein-nonlocal, however, or theories in which the level of reality handled by QM is underlain by a deeper reality (even within the proper scope of QM), such as that of classical random fields (Khrennikov 2010, Plotnitsky and Khrennikov 2015). The so-called superdeterminism is another realist view, which explains away the complexities discussed here by denying an independent decision of performing one or the other EPR measurements, a possibility central to Bohr's view and defining complementarity (e.g., 't Hooft 2001, 2018). These views, however, can be put aside here, because they either allow for Einstein-nonlocality or, if they claim Einstein-locality, they leave no space for the concept of quantum nonlocality in terms of predictions at a distance, because in all of these arguments, events are, as against the RWR view, always assumed to be connected by continuous and causal physical processes.

Bohr's interpretation in his reply was somewhat different from his ultimate interpretation (developed following his exchange with EPR), which no longer allowed for any assignment of elements of reality to quantum objects—before, at the time, or after measurement—because, as discussed in Section 3, the measured value of the quantum stratum was of the measuring instrument after the instrument had interacted with the object, a value "amplified" to the corresponding classical variable associated with the observed part of the instrument. This view emerged later, in following and under an impact of the Bohr-EPR exchange. In his reply, this assignment is possible *at the time* of measurement, or even, at least in responding to EPR's paper, on the basis of a prediction, but with crucial qualifications explained below.[34] However, given that this less radical view still allowed him to argue that quantum objects or their behavior cannot be considered independently, the essential logic of his reply, especially its argument for the irreducible role of measuring instruments and complementarity in considering quantum phenomena, could be presented in Bohr's later terms as well. Bohr argued that, because of this joint role, quantum phenomena, including of the EPR type, disallow EPR's conception of physical reality and the corresponding criterion of physical reality they introduce, or at least, the unqualified *way* in which the criterion was used by EPR. It is true that Bohr only argued for (along with Einstein-locality) the Bohr-completeness of quantum mechanics, rather than for its Einstein-completeness. QM, again, expressly is not Einstein-complete in Bohr's interpretation by virtue of its RWR-type character, because it does not offer a representation of the objects and processes responsible for quantum phenomena. However, this is all Bohr needed to do in countering EPR's argument. EPR did not contend that QM was not Einstein-complete, but rather that it was not even Bohr-complete because its predictions were not exhaustive, unless QM or quantum phenomena themselves allowed for Einstein-nonlocality. In his later communications, Einstein acknowledged that, if the statistical predictions of QM exhaust the possibilities of observation, thus making it Bohr-complete, one can also see it as Einstein-local (Born 2005, pp. 155, 205). As noted earlier, this still could not satisfy Einstein, because he saw Einstein-completeness as necessary for a fundamental theory (e.g., Born 2005, pp. 155, 166-170, 205; Einstein 1949a, p. 81).

It would extend the paper beyond its scope to give justice to EPR's argument and Bohr's reply. I shall only offer a sketch of the exchange and then proceed to EPR complementarity, which is the instance of the concept that reflects some of its deepest aspects (Bohr 1935, p. 700).[35] The crux of the EPR argument is that the EPR (idealized) thought-experiment allows for predictions with certainty concerning quantum objects without physically interfering with them by means of measurement, and thus, in EPR's view, "without in any way disturbing the system," in accordance with their "criterion of reality": "*If, without in any way disturbing a system, we can predict with certainty (i.e., with the probability equal to unity) the value of a physical quantity, then there exists*

---

[34] Schrödinger adopts a similar view of quantum measurement in his presentation of the nonrealism of "the doctrine" in his cat-paradox paper, which he notes implies that the "pointer position" of the instrument "*is always reproduced within certain error limits when the process is immediately repeated (on the same object, which in the meantime must not be exposed to any additional influences*" (Schrödinger 1935, p. 158). In Bohr's ultimate view, a pointer position would not correspond to any value of the object or the quantum stratum of the instrument that interacted with the object, but only to the value of the corresponding variable associated with the observable part of the instrument, "amplified" to this classical level. Schrödinger, as I said, does not consider this type of measurement architecture. See note 25 above.

[35] I have considered Bohr's reply in detail previously (Plotnitsky 2009, pp. 237-312; Plotnitsky 2012, pp. 107-136; Plotnitsky 2016, pp. 136-154). The present discussion, however, modifies these treatments on several key points.

*an element of physical reality corresponding to this physical quantity*" (Einstein et al 1935, p. 138; EPR's emphasis). This possibility would seem to circumvent the irreducible role of measuring instruments in the constitution of all quantum phenomena, and thus both the uncertainty relations and complementarity. As Bohr argued in his reply, the application of this criterion in considering quantum phenomena, including of the EPR type, poses difficulties, underappreciated by EPR or Einstein in his subsequent communications, primarily by disregarding the role of measuring instruments, still crucial, even in considering predictions concerning distant quantum objects.

An EPR prediction concerning a quantum object, $S_2$, of the EPR pair ($S_1$, $S_2$), is enabled by performing a measurement on another quantum object, $S_1$, with which, $S_2$, has previously been in interaction, but from which it is spatially separated at the time of the measurement on $S_1$. Specifically, once $S_1$ and $S_2$, are separated, QM allows one to simultaneously assign both the *distance between* the two objects and *the sum of their momenta*, because the corresponding Hilbert-space operators *commute*. With these quantities in hand, by *measuring* either the position or, conversely, the momentum of $S_1$, one can *predict exactly* either the position or the momentum for $S_2$ without physically interfering with, "disturbing," $S_2$, which would, EPR assumed, imply that one can simultaneously assign to $S_2$ both quantities as elements of reality pertaining to $S_2$. "The authors [EPR]," Bohr said in his reply, "therefore want to ascribe an element of reality to each of the quantities represented by such variables. Since, moreover, it is a well-known feature of the present formalism of quantum mechanics that it is never possible, in the description of the state of a mechanical system, to attach definite values to both of two canonically conjugate variables, [EPR] consequently deem this formalism to be incomplete, and express the belief that a more satisfactory theory can be developed" (Bohr 1935, p. 696). It follows, then, if EPR are correct, that the formalism would not even be Bohr-complete, insofar as QM does not predict all that is possible to ascertain. The only alternative, as EPR saw it, would be the Einstein-nonlocal nature of quantum phenomena or QM, an argumentation discussed below (Einstein et al 1935, p.141). They disallowed this possibility, as did Bohr, although, as noted earlier, it has been assumed by some. EPR's reasoning would equally apply to the Bell-Bohm type of EPR measurements for discrete variables, say, complementary spin-direction measurements.

Bohr counterargued that the situation does not allow one to dispense with the role of measuring instruments, because this role entails limitations on the *types* of measuring arrangements used in determining the quantities in question, even if one does so in terms of predictions without performing a measurement on the object, $S_2$, concerning which these predictions are made (but only a measurement on another quantum object, $S_1$, with which the object in question previously interacted). These limitations result from "an influence *on the very conditions which define the possible types of predictions [by measurements on $S_1$, regarding the future behavior of the system [$S_2$]*" (Bohr 1935, p.700). It is disregarding this influence (which is not a physical influence on $S_2$!), as EPR do, that gives EPR's criterion of reality its "essential ambiguity" when applied to quantum phenomena, which ambiguity disables their argument. By contrast, exposing the irreducible nature of this influence, as defining our measurements of $S_1$ and the resulting predictions concerning $S_2$, and taking this influence and, in the first place, these conditions into account, allowed Bohr to argue that neither EPR's argument for the incompleteness of QM nor their reasoning concerning the Einstein-nonlocality of QM, as an alternative, could be sustained.

As indicated earlier, Bohr's thinking concerning the situation eventually led him to his ultimate interpretation, in which *only what has already occurred* determines any physical quantity considered, and *not what has been predicted* (even with certainty) and is yet to be confirmed by a measurement. Bohr's reply, however, assesses EPR's argument on their terms, whereby it is possible to assign certain properties to quantum objects under the constraints of the uncertainty relations, rather than in terms of his ultimate interpretation. Bohr also assumes, as do EPR, that this assignment is possible on the basis of a *prediction* with "probability equal to unity" rather than only a *measurement*, but, in contrast to EPR, only predictions that are *in principle verifiable*. This is, as will be seen, a crucial qualification, necessary in considering quantum phenomena but missed by EPR.

Thus, both EPR and Bohr assume that the EPR experiment for ($S_1$, $S_2$) can be set in two alternative ways so as to predict, with a probability equal to unity, either one or the other of the complementary measurable quantities for $S_2$ on the basis of measuring the corresponding quantities for $S_1$. Bohr thus did not question the EPR experiment itself, which he saw as perfectly legitimate and, its idealized nature notwithstanding, revealing the deeper essence of quantum phenomena, and which, as explained earlier, he eventually came to see as pertaining to and even (with simple additional arrangements) as possible to stage, as an idealized case, in any quantum measurement. Let us call this assumption "*assumption A*."

On the basis of this assumption, EPR infer that *both* of these quantities can be assigned to $S_2$, even though it is impossible to do so simultaneously (in view of the uncertainty relations for the corresponding measurements on $S_1$). This makes QM incomplete (under EPR's criterion) because it has no mechanism for this assignment, unless one allows for Einstein-nonlocality (Einstein et al, p. 141). Let us call this inference "*inference E*" (for Einstein).

Bohr argued that, while *assumption A* is legitimate, *inference E* is unsustainable because, as discussed earlier in considering any quantum measurement on the entanglement model, a realization of the two situations necessary for the respective assignment of these quantities would involve two incompatible (complementary) experimental arrangements and, thus, *two different quantum objects*, and two different EPR pairs to prepare them. There is no physical situation in which this joint assignment is ever possible for *the same object*, either simultaneously or separately. If one makes the EPR prediction, with probability equal to unity, for the second object, $S_{12}$, of a given EPR pair, ($S_{11}$, $S_{12}$), one would always need a different EPR pair ($S_{21}$, $S_{22}$) to get to make the measurement on $S_{21}$, in order to make an alternative EPR prediction concerning $S_{22}$. I designate this inference as "*inference B*" (for Bohr).

Nor is an identical assignment of the single quantity ever possible, or in any event, ever guaranteed, for two "identically" prepared *objects* in the way it can be in classical physics. This is, as explained earlier, because quantum experiments cannot be controlled so as to identically prepare quantum objects but only so as to identically prepare measuring instruments, because the behavior of the instruments (of their observable parts) is classical. The quantum strata of measuring instruments throughout which they interact with quantum objects do not affect these preparations but only the outcome of an actual measurement. On the other hand, this interaction is quantum and, hence, "uncontrollable," the point brought up by Bohr at two key junctures of his argument (Bohr 1935, pp. 697, 700). It follows that the outcomes of repeated, identically prepared, experiments, including those of the EPR type, will, in general, be different. This circumstance makes statistical considerations unavoidable in the EPR experiment, even though each prediction involved can be made with probability equal to unity, which prediction by itself, as discussed in Section 4, still does not guarantee that the quantity thus predicated is an "element of reality" in quantum physics. This is because one can always measure the complementary variable, which disables the assignment of the variable thus predicted, and there is, in contrast to classical physics, no way to measure both variables simultaneously, and thus no longer any basis for the assignment. This aspect of the situation does not appear to have been realized by EPR, whose *inference E* and their argumentation implicitly depends on the possibility of the identical repetition of a given EPR experiment, precluded by *inference B*.

One can diagrammatically represent the situation as follows. Let $X$ and $Y$ be two complementary variables, either continuous or discrete, in the Hilbert-space formalism ($XY - YX \neq 0$) and $x$ and $y$ the corresponding physical measurable quantities ($\Delta x \Delta y \approx h$); ($S_1$, $S_2$) is the EPR pair of quantum objects; and $p$ is the probability of prediction, via the wave function, $\Psi$. Then:

The EPR experiment (in EPR's and Einstein's view, which considers one EPR pair):

$S_1$ $\qquad\qquad\qquad\qquad\qquad\qquad\qquad$ $S_2$
$X_1$ $\qquad$ $\Psi_1$ (with $p = 1$) $\rightarrow$ $\qquad\qquad$ $X_2$
$Y_1$ $\qquad$ $\Psi_2$ (with $p = 1$) $\rightarrow$ $\qquad\qquad$ $Y_2$

The EPR experiment (in Bohr's view, according to which two EPR pairs are always required for two EPR predictions):

$S_{11}$ $\qquad\qquad\qquad\qquad\qquad\qquad\qquad$ $S_{12}$
$X_{11}$ $\qquad$ $\Psi_1$ (with $p = 1$) $\rightarrow$ $\qquad\qquad$ $X_{12}$
$S_{21}$ $\qquad\qquad\qquad\qquad\qquad\qquad\qquad$ $S_{22}$
$Y_{21}$ $\qquad$ $\Psi_2$ (with $p = 1$) $\rightarrow$ $\qquad\qquad$ $Y_{22}$

This diagram is that of a complementarity, which may be called the EPR complementarity. This complementarity can be described as follows. Once one type of measurement (say, that of variable $X$) is performed on $S_{11}$, enabling the corresponding prediction on $S_{12}$, we irrevocably cut ourselves off from any possibility of making the alternative, complementary, measurement (that of $Y$) on $S_{11}$ and, thus, from the possibility of ever predicting the second variable for S$_{12}$ (Bohr 1935, p. 700). There is simply no way to define that variable for S$_{12}$, except, by a measurement and thus by disturbing S$_{12}$, which, however, defeats the very purpose of EPR's argument. By prediction, this could only be done on $S_{22}$, which is to say, by preparing another EPR pair and performing a complementary measurement of $Y$ on $S_{21}$, which will irrevocably prevent us from establishing $X$ for $S_{22}$.

As noted earlier, stemming from "our freedom of handling the measuring instruments, characteristic of the very idea of experiment" in all physics, our "free choice" concerning what kind of experiment we want to perform is essential to complementarity (Bohr 1935, p. 699). However, as against classical physics or relativity, implementing our decision concerning what we want to do will allow us to make only certain types of predictions and exclude the possibility of certain other, *complementary*, types of predictions, which would require a different experimental arrangement and in fact a different object to be measured. In the EPR case, it is only possible to establish both complementary quantities for two EPR pairs, ($S_{11}$, $S_{12}$) and ($S_{21}$, $S_{22}$), and never for one, and if we had predicted the second quantity, instead of the first one, for $S_{12}$, it would not, in general, be the same as it is for $S_{22}$. This is a manifestation of quantum causality. The situation is inherently statistical: if we repeat the experiment for yet another identically prepared pair, say, ($S_{31}$, $S_{32}$), so as to make predictions concerning the position of $S_{32}$, we can make such a prediction exactly, but the outcome of the measurement on $S_{32}$, will not in general be the same as for $S_{12}$ or $S_{22}$. Once we performed both types of measurement for many pairs, we will have statistically correlated measurements, commonly manifested in spin measurements of the Bell-Bohm type. The preceding argument would clearly apply to them.

Bohr does not explain the situation in terms of two different objects and EPR pairs necessary in order to make the second EPR prediction. As, however, the discussion of quantum measurement in Sections 2 and 3 suggests, this is at least an implication of his argument, given his insistence in his reply and elsewhere that "in the problem in question we are not dealing with a *single* specified experimental arrangement, but are *referring* to two different, mutually exclusive, arrangements" (Bohr 1987, v. 2 p. 57, 60; Bohr 1935, p. 699). In view of this mutual exclusivity, due to the irreducible role of the measuring instruments, the second quantity in question cannot in principle be assigned to the *same quantum object*, *once one such quantity is assigned*. However, we can always make an alternative choice in selecting a measuring arrangement and thus measuring or predicting the other complementary variable in question, which is a defining aspect of complementarity. The joint assignment is not possible even if one accepts EPR's criterion of reality, whereby such an assignment is made on the basis of a prediction, *unless we add the context of measurement to this criterion*, which in effect is what Bohr suggested. It is not possible once an experiment enabling one to make the first prediction is performed, because the first object $S_1$ (using the notation corresponding to EPR's view of the experiment) or $S_{11}$ (using the notation corresponding to Bohr's view of the experiment) is no longer available. The simultaneous assignment of both is, again, precluded by the uncertainty relations, which is recognized by EPR. They aim to show that the uncertainty relations could be circumvented by arguing that both variables could in fact *be assigned* to a given quantum object at any moment of time, although only one of them could be actually measured or predicted. This leads them to reason that QM is incomplete (even Bohr-incomplete), or else nonlocal. Bohr counterargues that the uncertainty relations or complementarity, both defined by the irreducible role of measuring instrument in the constitution of all quantum phenomena, those of the EPR-type included, disallow one *ever* to simultaneously assign both quantities to or even simultaneously define them for any quantum object, even in the EPR case. Bohr concludes:

> From our point of view we now see that the wording of the above mentioned criterion of physical reality proposed by Einstein, Podolsky, and Rosen contains an ambiguity as regards the meaning of the expression "without in any way disturbing a system." Of course there is in a case like that just considered no question of a mechanical disturbance of the system under investigation [the second object of the EPR pair considered, concerning which we make the EPR prediction at a distance] during the last critical stage of the measuring procedure. But even at this stage there is essentially the question of *an influence on the very conditions which define the possible types of predictions regarding the future behavior of the system*. Since these conditions constitute an inherent element of the description of any phenomenon to which the term "physical reality" can be properly attached, we see that the argumentation of the mentioned authors does not justify their conclusion that quantum-mechanical description is essentially incomplete. On the contrary this description, as appears from the preceding discussion, may be characterized as a rational utilization of all possibilities of unambiguous interpretation of measurements, compatible with the finite [quantum] and uncontrollable interaction between the object and the measuring instruments in the field of quantum theory. In fact, it is only the mutual exclusion of any two experimental procedures, permitting the unambiguous definition of complementary physical quantities, which provides room for new physical laws the coexistence of which might at first sight appear irreconcilable with the basic principles of science. It is just this entirely new situation as regards the description of physical phenomena that the notion of *complementarity* aims at characterizing. (Bohr 1935, p. 700; Bohr's emphasis)[36]

---

[36] In this passage the term "system" is interchangeable with the term "quantum object," which is the case throughout the Bohr-EPR exchange.

This elaboration, especially Bohr's claim concerning "the essential ambiguity" of EPR's use of their criterion and specifically that this ambiguity pertains to the meaning of EPR's expression "without in any way disturbing a system," have posed difficulties for Bohr's readers. Bohr acknowledged these difficulties and the main reason for them, essentially the fact, defining the argument of this paper, that one deals here with the impossibility of unambiguously considering quantum objects and their independent behavior because of the irreducible role of measuring instrument in the constitution of quantum phenomena. As he said: "I am deeply aware of the inefficiency of expression which must have made it very difficult to bring out the essential ambiguity involved in a reference to physical attributes of objects when dealing with phenomena where no sharp distinction can be made between the behavior of the objects themselves and their interaction with the measuring instruments" (Bohr 1949, Bohr 1987, v. 2, p. 61).

However, the elaboration and Bohr's meaning in this particular clause, "an ambiguity as regards the meaning of the expression 'without in any way disturbing a system'" pose no special difficulties given the preceding analysis. Once one quantity in question is established (even on the basis of a prediction, in accordance with EPR's criterion of reality) for $S_{12}$, we cannot ever establish the second quantity involved without measuring and hence *disturbing* $S_{12}$. Only one of these quantities could be established for $S_{12}$ without disturbing it, but once it is established, never the other quantity without disturbing it. We can establish such an alternative quantity without disturbing it only for a different quantum object, $S_{22}$, via a different EPR pair ($S_{21}$, $S_{22}$), by a measurement of a complementary type on $S_{21}$. These two determinations cannot be coordinated so as to assume that both quantities could be associated with the same object of the same EPR pair. The coordination of such events can only be statistical. We cannot establish both quantities for the same system *without in any way disturbing it*. The only way to establish the second quantity for this system would be to perform a measurement on and thus disturb it, which, however, would erase the determination of the first quantity, if one assumes, as the EPR do, that it could be made on the basis of a prediction on the first object of the corresponding EPR pair. This point, as will be seen below, is also crucial for maintaining the Einstein-locality of quantum mechanics. Thus, the ambiguity in question indeed relates to the clause "without in any way disturbing the system," which, if one wants to apply this clause rigorously in the EPR situation, requires qualifications explained in the present analysis but not provided by EPR. These qualifications amount to the fact that both quantities in question are complementary and as such can never be predicted for the same system, which would disable not only EPR's argument concerning the incompleteness (Bohr-incompleteness) of QM, as just explained, but, as will be seen presently, also concerning their alternative claim, that of Einstein-nonlocality.

Before I consider this issue, I reiterate that the considerations just offered could be transferred, with a few easy adjustments, to Bohm's version of the EPR experiment and spin variables. In this case, too, there is the EPR complementarity insofar as any assignment of the alternative spin-related quantity to the same quantum objects becomes impossible, once one such quantity is assigned. An assignment of the other would require an alternative type of measurement, mutually exclusive with the first, on the first object of a given pair, and hence, at least, another fully identically behaving EPR-Bohm pair, which is, again, not possible or at least cannot be guaranteed. Nothing other than statistical correlations between such assignments is possible, which is consistent with the Bell-EPR correlations, which are statistical. The argument, to which I now turn, concerning the Einstein-locality of EPR predictions could be similarly transferred to discrete variables as well.

EPR acknowledged a possible loophole in their argument by admitting that they did not demonstrate that one could ever *simultaneously* ascertain both quantities in question for the same quantum object, such as $S_2$ in the EPR experiment, in the same location, either that of $S_1$ or $S_2$.[37] They, however, see this requirement as implying Einstein-nonlocality:

> One could object to this conclusion on the grounds that our criterion of reality is not sufficiently restrictive. Indeed, one would not arrive at our conclusion if one insisted that two or more physical quantities can be regarded as simultaneous elements of reality *only when they can be simultaneously measured or predicted*. On this point of view, since either one or the other, but not both simultaneously, of the quantities *P* and *Q* can be predicted, they are not simultaneously real. This makes the reality of *P* or *Q* depend upon the process of measurement carried out on the first system, which does not disturb the first system in any way. No reasonable definition of reality could be expected to permit this. (Einstein et al 1938, p. 141)

*Einstein-nonlocality* indeed follows if one assumes, as EPR do, that the measurement, say, of *P*, on $S_1$ *fixes the physical state itself* of $S_2$ by "a spooky *action* at a distance," rather than allows "a spooky *prediction* at a distance," or *quantum nonlocality*, by fixing *the possible conditions* of such a prediction. It follows, under EPR's assumption, that an alternative measurement of *Q* on $S_2$ would discontinuously change this fixed state, although EPR do not examine this last eventuality as such. Or, as Einstein argued on later occasions, one is left with a paradoxical situation insofar as (assuming that QM is complete) two mutually incompatible states could be assigned to the same distant quantum object or system, $S_2$, by a different "spooky action at a distance," defined by a different measurement performed on $S_1$ (e.g., Born 2005, p. 205).[38] This is why EPR contend that,

---

[37] As has been noted by several authors, Schrödinger arguably the first of them (Schrödinger 1935, p. 160), one could simultaneously make alternative (complementary) measurements on $S_1$, say, the position measurement, which determines its position, and on $S_2$, the momentum measurement, which determines its momentum, and thus simultaneously predict (ideally exactly) the second variable for each system, the momentum for $S_1$ and the position for $S_2$. This determination, however, is *not simultaneous* in the same location.

[38] Einstein does note on the same occasion that the paradox is eliminated if quantum mechanics is only a statistical theory of ensembles and not of individual events, because, in this case, no single measurement of a given variable on $S_1$ or, more accurately, $S_{1n}$ determines the value of the corresponding variable on $S_{2n}$ (Born 2005, p. 205). These statistics involve correlations between spatially separated events, which has important implications for the question of locality, still in accord with the argument given here. Thus, as N. D. Mermin shows, in considering (essentially, correlatively to Bell's theorem) the statistical data obtained in the Bohm-EPR type experiments for two spatially separated devices for spin directions, A and B: "[It] is wrong to apply to individual runs of the experiment the principle that what happens at A does not depend on how the switch is set at B. Many people want to concluded from this that what happens at A *does* depend on how the switch is set at B, which is disquieting in view of the absence of any connections between the detectors. The conclusion can be avoided, if one renounces the Strong Baseball Principle, maintaining that indeed what happens at A does not depend on how the switch in set at B, but that this [independence] is only to be understood in its statistical sense, and most emphatically cannot be applied to individual runs of the experiment. To me this alternative conclusion is every bit as wonderful as the assertion of mysterious [spooky] action at a distance. I find it quite exquisite that, setting quantum metaphysics entirely aside, one can demonstrate directly from the data and the assumption that there are no mysterious [spooky] actions at a distance, that there is no conceivable way consistently to apply the Baseball Principle [what happens at A does not depend on how the switch in set at B] to individual events" (Mermin 1994, p. 109). This difference is equivalent to, indeed *is*, the difference between the Einstein-nonlocality of spooky *action* at a distance and quantum nonlocality of spooky *predictions* at a distance, predictions still spooky because, in the RWR view, we don't know and possibly cannot conceive how they come about and why QM correctly predicts them. Naturally, we cannot predict these correlations correctly on the basis of the data observed in one detector: "There is no way to infer from the data at one detector how the switch was set in other. Regardless of what is going on in detector B, the data for a great many runs at detector A is simply a random string of R's [red signals] and G's [green signals]." We can only predict these data correctly by QM (for discrete variables), if we know both settings. If, however, somebody, unbeknown to us, will change the setting in one detector (analogously to measuring the complementarity variable instead of the predicted one on the original EPR thought

if QM *is* complete by their criterion, then the physical state of a system, $S_2$, could be *determined* by a measurement on a spatially separate system, $S_1$, in violation of the Einstein locality, while their criterion of *reality* no longer applies in its original form. If it is Einstein-local, their main argument, based on their criterion of reality, showed (they believed) that it is incomplete.

Einstein thought that Bohr accepted the alternative of Einstein-locality vs. completeness (Einstein-locality vs. Bohr-completeness), and retained completeness by allowing for Einstein-nonlocality. Einstein, however, misread Bohr's argument, which only allows for a spooky *prediction*, and *not action*, at a distance, something running contrary to Einstein's "own way of thinking," into which, as he himself said, he "translated" Bohr's argument: "translated into my own way of putting it" (Einstein 1949b, p. 681). Beginning with bypassing the role of quantum measurement and ending with seeing nonlocality as acceptable to Bohr in order to preserve the completeness of quantum mechanics as concerns individual quantum systems, Bohr's argument is lost in Einstein's "translation." This "translation" reads Einstein's own logic of the relationships between locality and completeness into Bohr's very different logic, which, as noted, "ensures the compatibility between [his] argument and all exigencies of relativity theory," and thus Einstein-locality (Bohr 1935, p. 701n; Plotnitsky 2009, pp. 245-247).

There is a difference between determining, *fixing*, the state of a physical object *by* a prediction and *possibly* establishing it *on the basis* of a prediction, "a prediction with probability equal to unity." In Bohr's counterargument in his reply to EPR, physical states of quantum objects cannot be seen as finally determined (even when we have predicted them exactly) unless either the actual measurement is made or the possibility of *verifying* the prediction is assured insofar as such a measurement could, in principle, be performed so as to yield the predicted value. This last requirement in turn becomes a necessary qualification of EPR's criterion of reality in the case of quantum phenomena. This is because, as discussed in Section 4, in considering any prediction with probability equal to unity in quantum physics, if one assumes the validity of EPR's criterion in its original (unrestricted) form, the measurement of the alternative quantity, $Q$, on $S_2$ would automatically disable any possible verification of the original prediction. It is crucial and is, as noted, central to complementarity that it is always possible to perform this alternative measurement. This is one of the reasons why the assumption of the *independent existence or reality* of quantum objects or something in nature so idealized becomes especially important for Bohr's analysis of the EPR experiment and of the question of locality in it. This independent existence of reality ensures the possibility of this measurement. However, once this alternative measurement is performed, the original prediction becomes meaningless as in principle unverifiable. This, again, implies that both quantities in question could never be experimentally ascertained for the same object and hence that QM could not be shown to be (Einstein) nonlocal by EPR's logic, any more than it can be shown to be (Bohr) incomplete by their logic. That does not of course eliminate that either might in principle be the case, but only disables EPR's argument on both counts.

---

experiment), our predictions will no longer correspond to what is actually observed, and there would be no way to know whether they were actually correct, unless of course statistically from repetitions of the experiment with the original settings of both detectors. Accordingly, there is no experimental basis to ascertain that the quantum objects considered can be assigned both "elements of reality," one found in one setting and the other in the other. As Mermin notes, both the EPR-Bohr exchange concerning the EPR experiment and the Bell theorem "can be stated quite clearly in terms of the device [reproducing quantum correlations]" that Mermin uses for his demonstration, with the original EPR experiment beyond "more intricate but conceptually similar" (Mermin 1994, pp. 90-91). This is indeed my reason for considering Mermin's analysis here. It shows that considering quantum correlations, which are well confirmed experiments, for discrete variables is conceptually the same as that for continuous variables in the EPR-Bohr exchange.

According to Bohr, EPR's logic is disabled by the nature of quantum phenomena, as defined by the irreducible role of measuring instruments in the constitution of these phenomena, and thus by the impossibility of considering the behavior of quantum objects independently of their interaction with these instruments, in other words, noncontextually. The application of EPR's criterion of reality becomes, Bohr argues, "ambiguous" by virtue of the lack of qualifications of this criterion required by these conditions, which is to say, by complementarity. While they do not speak in terms of complementarity, EPR's paper or Einstein's subsequent communications *in effect* aim to show that, given the data obtainable in quantum phenomena, it should be possible, assuming Einstein-locality, to circumvent both the uncertainty relation and complementarity. This would make QM, which is consistent with both incomplete, even Bohr-incomplete, insofar as "its predictions [did] not exhaust the possibilities of observation" (Bohr 1987, v. 2, p. 57) or Einstein-nonlocal. Bohr counterargues that both are uncircumventable, because of the uncircumventable role of measuring instruments in the constitution of quantum phenomena, a role that defines the "conditions [that] constitute an inherent element of the description of any [quantum] phenomenon to which the term 'physical reality' can be properly attached" (Bohr 1935, p. 700).

## 6. Conclusion

By bringing together the irreducible role of measuring instruments, the RWR view, discreteness of quantum phenomena, complementarity, entanglement, and *quantum nonlocality*, QM may be interpreted as an *Einstein-local* and yet also *complete* theory, admittedly only Bohr-complete, in that it predicts all that nature allows us to predict, as things stand now. As noted earlier, Einstein eventually acknowledged that, if the statistical predictions of QM exhaust the possibilities of observation, QM would be Bohr-complete, while also allowing for Einstein-locality. Einstein, however, still found this insufficient for a fundamental theory, which he required to be Einstein-complete. He saw "the belief that should offer an exhaustive description of individual phenomena," by only proving the statistical predictions concerning the outcome of repeated experiments as "logically possible without contradiction," but found it "so very contrary to [his] scientific instinct that [he could not] forego the search for a more complete conception." (Einstein 1936, p. 375). This instinct required such a theory to be Einstein-complete, as well as Einstein-local, and Einstein, again, believed QM "offers no useful point of departure for future development" toward such a theory (Einstein 1949a, p. 83). He might have been right in this last score: QM is neither Einstein-complete, at least in the RWR view, nor may it ever lead to an Einstein-complete and Einstein-local theory of the ultimate constitution of nature. If one allows for Einstein-nonlocality, QM could, again, be seen as Einstein-complete as well, something, however, than neither Einstein nor Bohr found acceptable, specifically insofar as relativity remained in place.

Perhaps, however, the question is not what we require from a fundamental theory, unless experimental evidence leads to such requirements (which was, however, not Einstein's motivation for his imperative), but what a fundamental theory, either one already in place or one we need to develop, requires from us. One of the things it may require from us is a change of our attitude toward problems, such as that of physical reality, that we confront. I would argue, given Bohr's customarily careful way of expressing his points, that his statement "a radical revision of our *attitude* toward the problem of physical reality" (Bohr 1935, p. 697; emphasis added) need not mean that one should necessarily adopt any particular concept of reality, even though Bohr did adopt an RWR-type concept of reality as against a realist one. More important is our attitude itself toward this problem or any problem we confront, such that of quantum nonlocality: we should not be bound by previously established views, no matter how ingrained or cherished, and be ready to change our ways of thinking, that of RWR-type, among them, if physics requires it.

**Acknowledgements**. I am grateful to G. Mauro D'Ariano, Christopher A. Fuchs, Gregg Jaeger, and Andrei Khrennikov for helpful discussions concerning the subjects considered in this paper.

### Appendix: The physics of models, from Hertz to Bohr and beyond

Bohr's claims that "in quantum mechanics, we are not dealing with an arbitrary renunciation of a more detailed analysis of atomic phenomena, but with a recognition that such an analysis is *in principle* excluded" [beyond a certain point] (Bohr 1987, v. 2, p. 62), and Bohr's argument, as considered in this paper, leading to it (an argument that it is necessary to consider in order to properly understand the meaning of this conclusion), would, I would argue, complicate Khrennikov's assessment of Bohr's interpretation of quantum phenomena and QM. This assessment is implicit in Khrennikov's articles in question here and is stated expressly in (Khrennikov 2020b, p. 16). According to Khrennikov, for Bohr, QM was only an experimental model (EM), rather than a theoretical model (TM) in Hertz's and Boltzman's sense (e.g., Hertz 1999, Boltzmann 1974). This assessment is made by way of explaining Khrennikov's reservations concerning the idea of "*spooky* predictions at a distance," proposed by the present author. Khrennikov accepts (and even praises) the concept of a *prediction* at a distance (Khrennikov 2020a, p. 15). But he rejects spookiness, because he prefers, if not an actual spacetime description of the ultimate nature of reality underlying quantum phenomena, at least a theoretical model (TM), in Hertz's and Bolzmann's sense. Such a model would, in addition to an experimental model, EM, provide a spacetime idealization or (mental) picture (*Bild*) of this reality, an idealization or picture that need not strictly represent this reality. An EM would merely predict the outcomes of our experiments, now strictly in accord with the reality phenomena, represented by the observed phenomena, classical or quantum. While only an idealized "picture" rather than strictly a representation of reality, such a TM would still be realist in the present definition, just as would be Kant's view of the ultimate nature of reality as defined by things-in-themselves, which is part of the genealogy of Hertz's and Boltzmann's understanding of models. As Khrennikov notes, "[For Plotnitsky], 'spookiness' of quantum predictions is due to the impossibility to create any space-time picture explaining correlations" or for that matter explaining how any quantum phenomena come about (Khrennikov 2020a, p. 15). This is true, moreover, no matter how idealized such a picture or model would be, except that, as explained above, in the present view, as well as that of Bohr, this "impossibility" is only claimed a) still to be *an idealization*, an idealization without any picture, of the ultimate reality considered and b) only assumed *as things stand now*. The second qualification applies even if one adopts the strongest form of the RWR views, according to which the nature of quantum objects and their behavior, or, again, the reality so idealized, is beyond all *conception, now or ever*, because either the data in question in QM or our theories of concerning these data may change, which would render even this strong view no longer viable. The necessity of the qualification "as things stand now" is more apparent if one assumes a weaker form of the RWR view, in which this nature of beyond knowledge or any *currently available conception*, rather than any possible conceptions, but the qualification itself applies in either case. Hence, in the present view, too, just as that of Bohr, as Khrennikov contends, constructing a consistent classical-like TM remains in principle possible in the future (Khrennikov 2020a). It is only "*in principle* excluded," as things stand now. Of course, it is also possible that such a TM, and the corresponding realist interpretation of QM, would in principle be acceptable to those who hold this (EPR-type) already exists somewhere, is introduced by somebody, but is not known to or not accepted (or not understood) by those who adopt RWR-type interpretations. As I have stressed throughout, RWR-interpretations, including that of Bohr or the one adopted here, are only assumed here to be *interpretations*. They are only claimed, by Bohr or here, to be logically consistent and in accord with the experimental data of quantum physics, and not in conflict with experimentally confirmed findings elsewhere, all of which would be required for any interpretation or any TM. If coupled to the formalism of QM, RWR-type interpretations may be seen as establishing forms of TMs that are neither realist nor, as a consequence, classical causal, as TMs considered by Khrennikov are.

Khrennikov's argument leading to his assessment of Bohr's view, an assessment which, I argue, needs to be qualified on these grounds, is as follows. Khrennikov says first:

> By criticizing Plotnitsky, I again … refer to the Hertz-Boltzmann … methodology of scientific theories. By this methodology, there are two levels of scientific representation of natural phenomena:
> • a theoretical model (TM)
> • an experimental model (EM)
> TM respects the universal *law of cause and effect*.
>
> EM provides consistent description and prediction for experimental data. Already in 19th century scientists understood well (at least Hertz and following him Boltzmann) that experimental data is statistical and its description and prediction are probabilistic. For them, it was clear that EM need not be causal (cf. with von Neumann [von Neumann 1932], acausality of quantum measurements). Of course, TM and EM have to be coupled, TM→EM. However, coupling is not rigid, TM is not rigidly coupled to experiment, TM is our mental image ("Bild") of physical phenomena, its main aim is to respect the law of cause and effect. In short, Hertz and Boltzmann by developing the "Bild-concept" were precursors of Bell with his attempt to introduce hidden variables in quantum theory. (Khrennikov 2020a, p. 15)

Khrennikov qualifies that "For Hertz and Boltzmann, TM was not about reality as it is, but just its mental 'Bild', consistent and respecting the law of cause and effect" (Khrennikov 2020a, p. 17). (Both Hertz and Boltzmann follow Kant here.) In fact, any such TM obeys the law of classical causality, as defined in the present paper, which is confirmed by Khrennikov's footnote: "[This law] states that for every effect there is a definite cause, likewise for every cause, there is a definite effect" (Khrennikov 2020a, p. 15, n. 8). In part for that reason. Schrödinger, whom Khrennikov rightly credits with his appeal to Hertz and Bolzmann (Khrennikov 2020a, p. 15), refers to this model as "the classical ideal." He also offers a nuanced discussion of this ideal in the first section, "The Physics of Models," of his cat-paradox paper, the title I borrow as part of my title of this Appendix (Schrödinger 1935, pp. 152-153). He also acknowledges the difficulty, even if not impossibility (for he shared Einstein's hope for a theory of quantum phenomena in accord with this ideal, a hope that, for both Einstein and Schrödinger received new support by EPR's paper), of maintaining this ideal in the case of quantum phenomena, as handled by QM (Schrödinger 1935, p. 153). One wonders, accordingly, how universal the law of (classical) causality is in view of QM or even already Bohr's 1913 atomic theory, both discovered after both Hertz and Boltzmann were dead, although Boltzmann, who died in 1906, witnessed the emergence of quantum theory. For the moment, the present author, too, assumes, on this point in agreement with Hertz and Boltzmann, or, to begin with Kant, that any kind of TM, causal or not (as noted above it is possible to think of a noncausal TM), is only an idealized model of the reality considered and not this reality itself. Besides, a representation or even a conception of this reality is not possible on the RWR view in any event.

The main issue here is, then, the possibility of coupling of TM of the classical type and EM in the case of quantum phenomena and possibly (by way of a realist and causal interpretation) QM, a coupling representing Einstein's and Schrödinger's desideratum and even imperative, and, it appears, Khrennikov's own. The only example of such a coupling that Khrennikov provides is "*prequantum classical statistical field theory*, the classical random field model," developed by him previously, and conceptually (although, by dealing with fields, not physically) analogous to that of classical statistical physics, as its name suggests (Khrennikov 2020a, pp. 15-16, n10). If one strictly accepts this classical type or ideal of TM, then QM as currently constituted may indeed only be EM. This is what Einstein argued, possibly having in mind a classical-like TM that would approximate physical reality more closely and would, as explained earlier, not only be classically causal but also a *deterministic* model of the (field) type in classical physics and then relativity. This appeal to determinism is perhaps why Khrennikov, adopting the statistical view or EM of QM, with the possibility of the corresponding causal TM in mind, says that Einstein "wanted too much from the map TM→EM" (Khrennikov 2020a, p. 16).[39] On the other hand, Bohr's position is, in my view, a more complex matter than that of viewing QM merely as an EM. According to Khrennikov:

> Coming back to the Einstein-Bohr debate, we can say that Einstein said that QM is not TM, Bohr replied that he did not see a problem, since he knows that QM is EM. It seems that Bohr did not reject a possibility to construct a consistent TM for QM (treated as EM), but he would not accept the Hertz-Boltzmann-Schrödinger viewpoint on the structure of a scientific theory. He considered such kind of activity as metaphysical and, hence, meaningless. In contrast, Einstein badly wanted TM for QM, but (as latter Bell) he wanted too much from the map TM→EM. ….
>
> Now, [I?] turn to Plotnitsky's spooky predictions. Generally, predictions of any EM are spooky, since it is not EM's aim to present the causal picture of physical phenomena. The latter is done by TM. (Once again, QM≠TM, QM=EM.) So, I think that [the] terminology "spooky predictions" is misleading. (Khrennikov 2020a, p. 16-17).

I am not sure in what sense the "terminology 'spooky predictions' is misleading." Is it because a classical causal TM is possible? But, as I explained above, I do not deny this possibility either, anymore that Bohr does. I only say that, *as things stands now*, these predictions, moreover, *in the RWR-type interpretation* I adopted (or that of Bohr), are spooky because the correlations thus predicted cannot be understood in terms of a causal space-type behavior of quantum objects, which is, in this interpretation, beyond any representation or even conception, and thus beyond any classical TM. There is nothing spooky or mysterious about quantum correlations in terms of their mathematical predictions. They may, for example, be seen, as by Khrennikov, in terms of "*probability update on the basis of the quantum calculus*," even though this may still need an interpretation of the nature of our predictions, for example, either in statistical terms, as in Khrennikov, or in QBist (Bayesian) terms (e.g., Fuchs et al 2014). The physics and epistemology of quantum correlations—that is, why they are possible or why it may not be possible to know or even conceive of why they are possible, thus making them "mysterious" or "spooky"—is, however, a different matter. These are the physics and epistemology that are my concern and my main reasons for arguing for their "spooky" nature, again, using the term by way of juxtaposing spooky predictions at a distance to Einstein's spooky action at a distance. As I said, other terms are possible. What is important is this *concept* itself, defined by both the distant nature of these correlations and the impossibility of representing or conceiving of how they come about.

---

[39] On the other hand, Einstein's realism was not a naïve realism, given his insistence on the fundamental and indeed irreducible role of concepts in any physical theory or model, even an EM (e.g., Plotnitsky 2016, pp. 42-43). In fact, if anything, I would argue that Hertz in *The Principle of Mechanics* (Hertz 1899) might have been more of a naïve realist than Einstein, albeit not entirely naïve either. This is, however, a separate subject.

While, then, it is true that Bohr "did not reject a possibility to construct a consistent [classical-like] TM for QM," he also thought that such a TM may *not be possible*, and even believed that it was unlikely, if again, not strictly impossible, as things stand now, that is, insofar as QM and the experimental data it deals with are in place, as they still are. Hence, as I argue in this paper and as Khrennikov appears to suggest, the Bohr-Einstein debate was about whether such a classical-like TM of the ultimate constitution of nature is possible. Einstein's view was that it *should be* possible, while that of Bohr was that it *might not* be possible, which is, however, not the same as to argue that it will *never be possible*. While Einstein thought in terms of a deterministic TM→EM coupling, the same alternative would obtain for a probabilistic or statistical TM→EM coupling in the case of either QM or whatever theory might replace it, and Bohr's position would be the same as well. Einstein, as I said, also believed that QM was unlikely to be a right point of departure for such a deterministic future TM, which would predict the individual behavior of the ultimate constituents of nature. Or, again, while Einstein realized that there could not have been a full guarantee that such a model will even be developed, he strongly believed it could be, and he never stopped searching for a way toward it. As I said, the question remains open and the debate concerning it is as alive as ever, as Khrennikov's argument testifies.

While, then, it is true that for Bohr QM was not, or did not have, a classically causal or realist TM, it would, in my view, be difficult to claim that Bohr saw QM as a merely EM. It would, in my view, be also difficult to justify, on the basis of Bohr's writings, including his correspondence, the claim that Bohr "considered such kind of activity [or developing TMs for physical theories] as metaphysical and, hence, meaningless." Bohr, as I argue in this paper and related works cited here, was deeply concerned with the ultimate nature of reality and, given that he clearly realized that we can only have an idealization of this nature (which may not correspond to how things really are as concerns even our interactions with nature, primarily at stake in his view of quantum theory, let alone nature itself), what kind of idealization of this reality or the corresponding theory is possible. But he came to the conclusion, that, as things stand now, assuming, on RWR lines of thought, that assigning a representation or even conception the ultimate reality, idealized in terms of quantum objects and behavior, is "*in principle* excluded" leads to a viable *interpretation* of quantum phenomena and quantum mechanics. This view is, again, still an idealization and not the ultimate reality at stake in quantum physics, which makes this reality inconceivable even as inconceivable. On the other hand, one could see QM cum an RWR-type interpretation, as an (idealized) TM of a different, RWR, type (e.g., Plotnitsky 2016, pp. 5-6). In any event, this is the present interpretation of Bohr's interpretation, which is sometimes interpreted in stronger terms by bypassing the qualification "as things stand now," which is, in my view, as important for Bohr's argument as it is for my argument here.

I would like to add a historical comment, responding to Khrennikov's remarks accompanying his argumentation just considered. According to Khrennikov: "Unfortunately, creators of QM were not aware about the 'Bild-concept' of Hertz-Boltzmann (or just ignored it? Schrödinger tried to appeal to it, but his message was completely ignored). … It is surprising that [current] philosophers (who really read a lot) are not aware about [of?] the works of Hertz and Boltzmann" (Khrennikov 2020a, p. 16, p. 17, p. 16 p.16n). I doubt that either suggestion is true, especially, the first one, although it is true that some and even most of the creators of QM would *reject* this type of concept, which, however, is not the same as ignoring it. Hertz's and Boltzmann's arguments concerning the subject or at least their key ideas were almost certainly familiar to most founders of QM, rather than only Schrödinger, either directly or via mediation by others. As Khrennikov says, "The situation is really strange. Everything happened nearby [in] Germany, all could read in German, and Hertz and Boltzmann were really famous," after asking "this is the good question to philosophers of science: Did Bohr and Einstein (as well as [,] say [,] Heisenberg and von Neumann) know about the works of Hertz and Boltzmann?" (Khrennikov 2020a, p. 16, p. 17, p. 16, p.16n.) The specifics many indeed need a good historian of science to be established, but it is hardly in doubt that the key ideas in question were in circulation and were familiar to most founding figures of QM, for one thing, because these ideas where questioned and challenged, and sometimes renounced, by them in view of quantum phenomena and QM. One might argue that Hertz's early death made his more specific ideas (some quite remarkable, including his understanding of the symbolic nature of physical theories and models or his use *n*-dimensional space) in *The Principles of Mechanics* from having a more immediate impact. Its basic ideas concerning models were known, however, even apart from Boltzmann's and then Schrödinger's contribution to their currency, and these types of ideas had other sources, philosophically, again, traceable at least to Kant. D. Hilbert, in particular, was well aware of them, as he was of those of Boltzmann, and Hilbert lectured on both Hertz and Boltzmann in Göttingen (e.g., Corry 2004). Both Born and Einstein were certainly familiar with their work. Heisenberg and Pauli, both of whom spent years in Göttingen, were likely to have some sense of Hertz's and Bolzmann's views as well. Heisenberg's and Pauli's opposing views concerning realism and, especially, classical causality needs not suggest otherwise, quite the contrary. On the other hand, it is true that neither engaged with Hertz's and Bolzmann's arguments, possibly because they saw those arguments as unworkable in quantum theory.

Finally, although establishing such connections is often complicated in Bohr's case, Bohr was certainly familiar with Boltzmann's ideas in question, and the connections between Bohr's thinking and some of Hertz's ideas have been suggested as well (e.g., Chevalley 1993, p. 48). I would even argue that Bohr's argument concerning the impossibility of visualization and more specifically pictorial visualization of quantum objects and behavior, including in the context of the EPR experiment, "suited to emphasize how far, in quantum theory, we are beyond the reach of *pictorial* visualization" (Bohr 1987, v. 2, p. 59; emphasis added), may well have been made with the concept of "*Bild*" in mind. In both cases, moreover, that of Hertz and Boltzmann (or that of Kant) and that of Bohr, the idea of *Bild* and visualization extended to the capacity of our phenomenal or intuitive representation in general, *Anschaulichkeit,* the German word, used by Bohr sometimes. It is closer to the idea of something that appears, *shows itself*, to our thought, which was how Kant defined "phenomena," vs. "noumena" or objects, which are things-in-themselves (material or mental) as they are. As I also argue, in Bohr's and the present view, but, as noted, not that of Heisenberg's in his later thinking, the ultimate reality responsible for quantum phenomena is also beyond an idealized mathematical representation.

# References

Arndt, M., Nairz, O., Voss-Andreae, J., Keller, C., van der Zouw, G., Zeilinger, A.: Wave particle duality of C60, *Nature* **401**, 680–682 (1999)

Aspect, A., Dalibard, J., Roger, G.: Experimental test of Bell's inequalities using time varying analyzers, Phys. Rev. Lett. **49**, 1804-1807 (1982)

Barbour, J. B.: *The End of Time: The Next Revolution in Physics*, Oxford University Press, Oxford (1999)

Bell J.S.: *Speakable and Unspeakable in Quantum Mechanics*, Cambridge, UK: Cambridge University Press (2004)

Bohr, N.: On the constitution of atoms and Molecules (Part 1), Philosophical Magazine **26** (151), 1-25 (1913)

Bohr, N.: The quantum postulate and the recent development in atomic theory, Nature **121** (827), 580-590 (1928)

Bohr, N.: *Atomic Theory and the Description of Nature*, Cambridge, UK: Cambridge University Press (1934)

Bohr, N.: Can quantum-mechanical description of physical reality be considered complete? Phys. Rev. **48**, 696-702 (1935)

Bohr, N.: Causality and complementarity. In Faye, J., Folse, H. J., (eds.) *The Philosophical Writings of Niels Bohr, Volume 4: Causality and Complementarity, Supplementary Papers*, Woodbridge CT, Ox Bow Press, CT 1994, 83-91 (1937)

Bohr, N.: The causality problem in atomic physics. In Faye, J., Folse, H. J., (eds.) *The Philosophical Writings of Niels Bohr, Volume 4: Causality and Complementarity, Supplementary Papers*, Woodbridge CT, Ox Bow Press, 1987, 94-121 (1938)

Bohr, N.: *The Philosophical Writings of Niels Bohr*, 3 vols, Woodbridge CT, Ox Bow Press (1987)

Born, M.: *The Einstein-Born Letters* (tr. Born, I.), New York, Walker (2005)

Born, M., Jordan, P.: Zur Quantenmechanik, Z. für Physik **34**, 858 (1925)

Boltzmann, L.: On the development of the methods of theoretical physics in recent times. In McGuinness, B. (ed.) Theoretical Physics and Philosophical Problems. Vienna circle collection, Vol. 5, Springer, Dordrecht, 13-32 (1974)

Brukner, C.: Quantum causality, Nature Physics **10**, 259- 263 (2014)

Brunner N., Gühne O., Huber, M (eds): Special issue on 50 years of Bell's theorem, J. Phys. A **42**, 424024 (2014)

Cabello, A: Interpretations of quantum theory: a map of madness. In Lombardi, O., Fortin, S., Holik, F., López, C. (eds) *What is Quantum Information?* Cambridge University Press, Cambridge, 138-144 (2017)

Chevalley, C. Niels Bohr's Words and the Atlantis of Kantianism. In Faye, J., Folse, H. (eds.) *Niels Bohr and Contemporary Philosophy*, Springer, Berlin, 33-55 (1993)

Corry, L.: *David Hilbert and Axiomatization of Physics (1898-2018)*, Springer, Berlin (2004)

Cushing J. T., McMullin E. (eds): *Philosophical Consequences of Quantum Theory: Reflections on Bell's Theorem*, Notre Dame, IN: Notre Dame University Press (1989)

D'Ariano, G. M.: Physics without physics, Int. J. Theo. Phys. **56** (1), 97-38 (2017)

D'Ariano, G. M.: Causality re-established. Phil. Trans. R. Soc. A 376: 20170313, http://dx.doi.org/10.1098/rsta.2017.0313 (2018)

D'Ariano, G.M., Chiribella, G., Perinotti, P.: *Quantum Theory from First Principles: An Informational Approach*, Cambridge UK, Cambridge University Press (2017)

Dirac, P. A. M.: *The Principles of Quantum Mechanics*, Oxford UK, Clarendon, rpt. 1995 (1958)

Einstein, A.: Physics and reality, Journal of the Franklin Institute, 221, 349–382 (1936)

Einstein, A.: *Autobiographical Notes* (tr. Schillp, P. A.), La Salle, IL: Open Court, 1991 (1949a)

Einstein, A.: Remarks to the essays appearing in this collective volume. In Schillp, P. A. (ed), *Albert Einstein: Philosopher–Scientist,* New York, Tudor, 663-688 (1949b)

Einstein, A., Podolsky, B., and Rosen, N.: Can quantum-mechanical description of physical reality be considered complete? In Wheeler, J. A., Zurek, W.H. (eds), *Quantum Theory and Measurement*, Princeton University Press, Princeton NJ, 1983, 138-141 (1935)

Ellis J., Amati D. (eds): *Quantum Reflections*, Cambridge, UK: Cambridge University Press (2000)

Fuchs, C. A., Mermin, N. D., Schack, R.: An introduction to QBism with an application to the locality of quantum mechanics, Am. J. Phys. **82**, 749. http://dx.doi. org/10.1119/1.4874855 (2014)

Gomes, H. de A.: Back to Parmenides, arXiv: 1603.01574 (2016)

Greenberger, D. M., Horne, M. A., and Zeilinger, A.: Going beyond Bell's theorem (1989). In Kafatos, M. (ed.) *Bell's Theorem, Quantum Theory and Conceptions of the Universe*, Kluwer, Dordrecht, 69-72 (1989)

Greenberger, D. M., Horne, M.A., Shimony, A., and Zeilinger, A.: *Bell's Theorem Without Inequalities*, Amer. J. Phys **58**, 1131-1142 (1990)

Hardy, L.: Nonlocality for two particles without inequalities for almost all entangled states, Phys. Rev. Lett. 71, 1665-1668 (1993)

Hardy, L.: Quantum mechanics from five reasonable axioms," arXiv: quant-ph/0101012 (2001)

Hardy, L.: A formalism-local framework for general probabilistic theories, including quantum theory. arXiv:1005.5164 [quant-ph] (2010)

Heisenberg, W.: Quantum-theoretical re-interpretation of kinematical and mechanical relations. In Van der Waerden, B.L. (ed.) *Sources of Quantum Mechanics*, New York, Dover, 1968, 261-272 (1925)

Heisenberg, W.: The physical content of quantum kinematics and mechanics. In Wheeler, J. A., Zurek, W. H. (eds.) *Quantum Theory and Measurement*, Princeton University Press, Princeton, NJ, 1983, 62–86 (1927)

Heisenberg, W.: *The Physical Principles of the Quantum Theory*, tr. Eckhart, K., Hoyt, F.C., New York, Dover, rpt. 1949 (1930)

Heisenberg, W.: *Physics and Philosophy: The Revolution in Modern Science,* Harper & Row, New York (1962)

Hertz, H. Hertz, H. (1899). The Principles of Mechanics: Presented in a New Form, Macmillan, London (1899)

Jaeger, G.: *Quantum Objects: Non-local Correlations, Causality and Objective Indefiniteness in the Quantum World*, New York: Springer (2013)

Jaeger, G.: Quantum contextuality in the Copenhagen approach, Phil. Trans. R. Soc. A, https://doi.org/10.1098/rsta.2019.0025 (2019)

Kant, I.: *Critique of Pure Reason*, tr. Guyer, P., Wood, A.W., Cambridge, UK, Cambridge University Press, (1997)

Khrennikov, A.: Demystification of quantum entanglement, arXiv: 0905.4791v3 [physics.gen-ph] (2010)

Khrennikov, A.: Quantum probabilities and violation of CHSH-inequality from classical random signals and threshold type detection scheme, Prog. Theor. Phys. **128**, 31–58. (doi:10.1143/PTP.128.31) (2012)

Khrennikov, A.: Get rid of nonlocality from quantum physics, Entrop*y* **21**(8), 806 (2019a)

Khrennikov, A.: Quantum versus classical entanglement: eliminating the issue of quantum nonlocality. https://arxiv.org/abs/1909.00267 (2019b)

Khrennikov, A.: Echoing the recent Google success: Foundational roots of quantum supremacy, arXiv: 1911.10337 (2019c)

Khrennikov, A.: Two faced Janus of quantum nonlocality, arXiv 2002.01977v1 [quantum-ph] (2020a)

Khrennikov, A.: Quantum versus classical entanglement: eliminating the issue of quantum nonlocality, Found. Phys. https://doi.org/10.1007/s10701-020-00319-7 (2020b)

Kupczynski, M.: Closing the Door on Quantum Nonlocality, Entropy **20**, 877; doi:10.3390/e20110877 (2018)

Ladyman, J.: Structural realism. The Stanford encyclopedia of philosophy (Winter 2016 edition), E.N. Zalta, E. N. (ed), https://plato.stanford.edu/archives/win2016/entries/structural-realism/ (2016).

Mehra, J., and Rechenberg, H.: *The Historical Development of Quantum Theory*, 6 vols. Berlin, Springer (2001)

Mermin, N. D.: *Boojums All the Way Through: Communicating Science in a Prosaic Age*, Cambridge, UK, Cambridge University Press (1990)

Mermin, N. D.: *Why Quark Rhymes with Pork: And Other Scientific Diversions*. Cambridge, UK, Cambridge University Pres (2016)

Plotnitsky, A.: *Epistemology and Probability: Bohr, Heisenberg, Schrödinger and the Nature of Quantum-Theoretical Thinking*. Springer, New York (2009)

Plotnitsky, A.: 'Dark materials to create more worlds': On causality in classical physics, quantum physics, and nanophysics, Journal of Computational and Theoretical Nanoscience **8** (6), 983-997 (2011)

Plotnitsky, A.: *Niels Bohr and Complementarity: An Introduction*, New York, Springer (2012)

Plotnitsky, A.: A matter of principle: The principles of quantum theory, Dirac's equation, and quantum information," Fond. Phys. **45** (10), 1222-1268 (2014)

Plotnitsky, A.: *The Principles of Quantum Theory, from Planck's Quanta to the Higgs Boson: The Nature of Quantum Reality and the Spirit of Copenhagen*. New York, Springer/Nature (2016)

Plotnitsky, A.: Comprehending the Connection of Things: Bernhard Riemann and the Architecture of Mathematical Concepts. In Li, J.,Yamada, S., Papadopoulos, A. (eds), *From Riemann to Differential Geometry and Relativity*, 329-363, Springer, Berlin (2017)

Plotnitsky, A.: "The Heisenberg Method": Geometry, Algebra, and Probability in Quantum Theory, Entropy **20**, 656; doi:10.3390/e20090656 (2018a)

Plotnitsky, A.: Structure without Law: From Heisenberg's Matrix Mechanics to Structural Nonrealism, Mind and Matter 16 (1) 59-96 (2018b)

Plotnitsky, A.: Transitions without connections: quantum states, from Bohr and Heisenberg to quantum information theory, Eur. Phys. J. https://doi.org/10.1140/epjst/e2018-800082-6 (2019a)

Plotnitsky, A.: Spooky predictions at a distance: reality, complementarity and contextuality in quantum theory, Phil. Trans. R. Soc. A 377: 20190089. http://dx.doi.org/10.1098/rsta.2019.0089 (2019b)

Plotnitsky, A.: On the concept of curve: geometry and algebra, from mathematical modernity to mathematical modernism. In Papadopoulos, A., Dani, S., *Geometry in History,* Berlin, Springer/Nature (2019c)

Plotnitsky, A.: "Without in any way disturbing the system:" Illuminating the issue of quantum nonlocality, arXiv:1912.03842 [quant-ph] (2019d)

Plotnitsky, A., Khrennikov, A.: Reality without Realism: On the Ontological and Epistemological Architecture of Quantum Mechanics. Found. Phys. **25** (10), 1269–1300 (2015)

Rovelli, C.: Relational Quantum Mechanics, Int. J. Theo. Phys. **35** (8) 1637-78 (1996)

Scully, M. O., Drühl, K. Quantum eraser: A proposed photon correlation experiment concerning observation and delayed choice in quantum mechanics, Phys. Rev. A **25**, 2208-2213 (1982)

Schrödinger, E.: The present situation in quantum mechanics. In Wheeler, J.A., Zurek, W.H. (eds), *Quantum Theory and Measurement*, Princeton University Press, Princeton, 1983, 152-167 (1935a)

Smolin, L., *Einstein's Unfinished Revolution: The Search for What Lies Beyond the Quantum*, New York: Penguin (2018)

Sorkin, R., D., Spacetime and causal sets. In D'Olivo, J. C., Nahmad-Achar, E., Rosenbaum, N., Rayn, M. P. Jr., Urrutia, L. F., Zertuche, F. (eds) *Relativity and Gravitation: Classical and Quantum*, Singapore: World Scientific (1991)

Spekkens, R. W.: Evidence for the epistemic view of quantum states: A toy theory, Phys. Rev. A **75.3** 032110 (2007)

Spekkens, R. W.: Quasi-quantization: classical statistical theories with an epistemic restriction. In: Chiribella, G., Spekkens, R. W. (eds) *Quantum Theory: Informational Foundations and Foils*, New York, Springer/Nature, 83-136 (2016)

'Hooft, G. Quantum mechanics and determinism. In Frampton, P., Ng, J. (eds) *Particles, strings, and cosmology*, Rinton Press, Princeton, NJ, 275–285 (2001)

't Hooft, G.: Time, the arrow of time, and quantum mechanics, arXiv: 1804.01383 (2018)

Von Neumann, J.: *Mathematical Foundations of Quantum Mechanics*, tr. R. T. Beyer, Princeton University Press, rpt. 1983, Princeton, NJ (1932)

Werner, R. F.: Comment on 'What Bell did', J. Phys. **A** 47: 424011(2014)

Wheeler, J. A.: Law without law. In Wheeler J.A., Zurek, W.H. (eds) *Quantum Theory and Measurement*, Princeton University Press, Princeton, NJ, 182-216 (1983)

Wheeler, J.A.: Information, physics, quantum: the search for links. In Zurek, W.H. (ed) *Complexity, Entropy, and the Physics of Information*. Addison-Wesley, Redwood City, CA, 3-28 (1990)

Wigner, E.: On unitary representations of the inhomogeneous Lorentz group, Ann. Math. 40, 149-204 (1939)

Wittgenstein, L.: *Tractatus Logico-Philosophicus* (tr. C. K. Ogden) London: Routledge, rpt. 1924 (1985)

Zeilinger, A., G. Weihs, T. Jennewein, and M. Aspelmeyer, Happy centenary, photon, Nature **433,** 230-238 (2005)